\documentclass[preprint,12pt]{elsarticle}

\usepackage{amssymb}
\usepackage{amsmath}
\usepackage{amsthm}
\usepackage[capitalize]{cleveref}
\usepackage{bm}
\usepackage{mathtools}
\usepackage{suffix}
\usepackage{ dsfont }
\usepackage{ bbold }
\usepackage{eufrak}
\usepackage[utf8]{inputenc}
\usepackage[T1]{fontenc}

\def\ta{t^a}
\def\tac{t^{a*}}
\def\1{1}

\newtheorem*{remark}{Remark}

\journal{Journal of Computational Physics}

\begin{document}

\begin{frontmatter}

\title{Trace preserving quantum dynamics using a novel reparametrization-neutral summation-by-parts difference operator}

\author[1]{Oskar $\rm \mathring{A}$lund}
\ead{oskar.alund@liu.se}

\author[2]{Yukinao Akamatsu}
\ead{akamatsu@kern.phys.sci.osaka-u.ac.jp}

\author[1]{Fredrik Laur\'en}

\ead{fredrik.lauren@liu.se}

\author[2]{Takahiro Miura}
\ead{miura@kern.phys.sci.osaka-u.ac.jp}

\author[1,3]{Jan Nordstr\"om}

\ead{jan.nordstrom@liu.se}

\author[4]{Alexander Rothkopf}

\ead{alexander.rothkopf@uis.no}

\address[1]{Department of Mathematics, Computational Mathematics, Link\"oping University, SE-581 83 Link\"oping, Sweden}
\address[2]{Department of Physics, Osaka University, Toyonaka, Osaka, 560-0043, Japan}
\address[3]{Department of Mathematics and Applied Mathematics, University of Johannesburg,P.O. Box 524, Auckland Park 2006, South Africa}
\address[4]{Faculty of Science and Technology, University of Stavanger, 4021 Stavanger, Norway}

\begin{abstract}
We develop a novel numerical scheme for the simulation of dissipative quantum dynamics following from two-body Lindblad master equations. All defining continuum properties of the Lindblad dynamics, hermiticity, positivity and in particular trace conservation of the evolved density matrix are preserved. The central ingredient is a new spatial difference operator, which not only fulfils the summation by parts (SBP) property but also implements a continuum reparametrization property. Using the time evolution of a heavy-quark anti-quark bound state in a hot thermal medium as an explicit example, we show how the reparametrization neutral summation-by-parts (RN-SBP) operator preserves the continuum properties of the theory.
\end{abstract}

\begin{keyword}
time integration\sep initial value problems\sep initial boundary value problems\sep dissipative systems\sep open quantum systems\sep summation-by-parts operators\sep mimetic operator
\end{keyword}

\end{frontmatter}


\section{Introduction}

The dynamical evolution of dissipative quantum systems constitutes the focus of a broad and active research community, ranging from quantum information science to condensed matter physics and even to high energy nuclear physics. Take the physics of impurities for example, where a heavy particle is introduced into a reservoir of light particles. The interactions of the impurity with the surrounding environment change its properties, leading to a wealth of phenomenologically relevant quantum phenomena, as well as decoherence. In the context of condensed matter physics such impurities have been discussed as Bose polarons (see e.g.~\cite{lampo_bose_2017}). On the other hand in the context of heavy-ion collisions it is the behavior of heavy quark-antiquark pairs, so called heavy quarkonium and their interaction with the hot collisions remnants that are of interest (for a recent review see \cite{Rothkopf:2019ipj}). Investigating the physics of the small subsystem inevitably leads one to consider dissipative dynamics and to the framework of open quantum systems (for an excellent introduction see \cite{BRE02}). 

It is common to consider a quantum system, which consists of a small subsystem (S), immersed in a large environment (E). Its state is encoded in the eponymous state-vectors $|\psi\rangle$ that are elements of the full Hilbert space ${\cal H}$. The subsystem and environment degrees of freedom, described by $|\psi^{\rm S}\rangle$ and $|\psi^{\rm E}\rangle$ respectively, can be found in separate subspaces ${\cal H}={\cal H}_S \otimes {\cal H}_E $. The symbol $\otimes$ denotes the direct product. The overall system is closed and its physics described by the hermitian Hamilton operator $\hat H^{\rm tot}$ acting on state vectors in the full Hilbert space. It can be decomposed into the following sum
\begin{align}
\hat H^{\rm tot}=\hat H_{\rm S}\otimes I_{\rm E} + I_{\rm S} \otimes \hat H_{\rm E} + \hat H_{\rm int}. \label{eq:Hamiltonian}
\end{align}
The first term $\hat H_{\rm S}$ refers to the Hamiltonian governing the dynamics of the subsystem. It leaves the environment unchanged, as indicated by the direct product with the identity operator $I_{\rm E}$ on the environment subspace of the full Hilbert space. $\hat H_{\rm E}$ describes the environment degrees of freedom and the interactions between the two are contained in $\hat H_{\rm int}$. The latter explicitly couples the subsystem and environment subspace of the full Hilbert space.

Many relevant properties of such a system are captured by its density matrix operator 
\begin{align}
\hat \rho^{\rm tot}=\sum_i p_i |\psi_i\rangle \langle \psi_i|,\label{eq:densitymat}
\end{align}
where we denote the adjoint of a state vector by $|\psi_i\rangle^\dagger=\langle \psi_i|$. 
The density matrix thus represents the outer product of all states realized in the system under consideration, weighted by their probability $p_i$. In the so-called Schr\"odinger picture $\hat \rho^{\rm tot}$ evolves according to the von-Neumann equation
\begin{align}
\frac{d}{dt}\hat \rho^{\rm tot}(t)=-i[\hat H^{\rm tot}(t),\hat \rho^{\rm tot}(t)],
\end{align}
where $[A,B]=AB-BA$ denotes the commutator. The hermiticity of $\hat H^{\rm tot}$ translates into unitary time evolution for the total density matrix $ \rho^{\rm tot}(t)=U(t,0)\rho^{\rm tot}(0)U^\dagger(t,0)$ implemented via the time evolution operator $U(t,0)={\cal T}{\rm exp}[-i\int dt H_{\rm int}(t)]$. Here the adjoint of the operator is denoted by the $(\dagger)$ symbol $U^\dagger(t,0)=U^{-1}(t,0)=U(0,t)$ and the symbol ${\cal T}$ refers to time ordering, relevant for explicitly time dependent Hamilton operators (for a more detailed exposition we refer the reader to \cite{sakurai2014modern}).

Measurable properties of the quantum system are encoded in hermitian operators ($\hat A^\dagger=\hat A$), so called observables. Their expectation value, representing the mean value of the associated physical property, obtained over repeated experiments, may be computed via trace over the density matrix  $\langle \hat A\rangle={\rm Tr}[\hat \rho \hat A]$. For observables related to the small subsystem, we now wish to simplify the description. Instead of having to carry out the trace over the full Hilbert space every time we compute an expectation value, we carry out the trace over the environment degrees of freedom in the full Hilbert space a priori. This leads us to the reduced density matrix
\begin{align}
    \hat \rho_S = {\rm Tr}_E[\hat \rho^{\rm tot}] = \sum_{n} \langle \psi^{\rm E}_n| \Big(\sum_i  p_i |\psi_i\rangle \langle \psi_i| \Big) |\psi^{\rm E}_n\rangle= \sum_l \tilde p_l |\psi_l^{\rm S}\rangle \langle \psi_l^{\rm S}|,\label{eq:reddenmat}
\end{align}
where we have expressed the trace as sum over the inner products with the n-th environment state vector $\langle \psi^{\rm E}_n| \hat \rho |\psi_n^{\rm E}\rangle$. Hence, in case that $\hat A= \hat A_{\rm S}\otimes I_{\rm E}$, we have $\langle \hat A \rangle(t) = {\rm Tr}[\hat \rho(t) \hat A] = {\rm Tr}_S[\hat \rho_S(t) \hat A_S]$. We thus need an evolution equation for the reduced density matrix $\hat \rho_S$.

In general, the time scales of the evolution in the subsystem and the environment may be of the same order and the time evolution of $\hat \rho_S$ remains genuinely non-Markovian, i.e. memory effects play a role. On the other hand if there exists a separation of timescales between the medium and the subsystem, i.e. if the subsystem takes longer to relax than the environment correlations take to decay, one may encounter approximate Markovian dynamics, in which memory effects can be neglected. In that case it has been shown that the most general equation of motion for $\hat \rho_S$ can be written in terms of the (Gorini–Kossakowski–Sudarshan)-Lindblad equation \cite{Gorini:1975nb,Lindblad:1975ef}
\begin{align}
\frac{d}{dt}\hat\rho_{\rm S}= -i[\tilde H_{\rm S},\hat\rho_{\rm S}]+\sum_k \gamma_k\Big( \hat L_k \hat\rho_{\rm S} \hat L^\dagger_k -\frac{1}{2} \hat L^\dagger_k \hat L_k\rho_{\rm S} - \frac{1}{2}\hat\rho_{\rm S} \hat L^\dagger_k \hat L_k\Big).\label{eq:Lindblad}
\end{align}
This equation describes the in-general non-unitary, i.e. dissipative, time evolution of the subsystem S through its reduced density matrix $\hat \rho_{\rm S}$. I.e. all operators here act only on states $|\psi_{\rm S}\rangle \in {\cal H}_{\rm S}$ in the subsystem subspace of the full Hilbert space. The influence of the environment E manifests itself in the presence of Lindblad operators $\hat L_k$, damping rates $\gamma_k$ and possible modifications of the subsystem Hamiltonian $\tilde H_S \neq \hat H_S$ (explicit examples of these quantities will be derived in \cref{sec:QQbarLindblad}). Formulating the dissipative dynamics in terms of a Lindblad equation is advantageous, as it can be proven that this equation preserves the main physical properties of the reduced density matrix, i.e. positivity, hermiticity and unitarity
\begin{align}
\langle \psi^{\rm S}_n | \rho_{\rm S} | \psi^{\rm S}_n \rangle >0, \forall n, \quad \rho_{\rm S}^\dagger = \rho_{\rm S}, \quad {\rm Tr}[\rho_{\rm S}]=1. \label{eq:Lindbprop}
\end{align}
The preservation of unit trace in particular is important, as the probability interpretation of the density matrix rests upon it. The reason is that understanding the energy, momentum and particle exchange between the subsystem and environment is of central interest in our study. Thus it is paramount to ensure that the formulation of the problem does not introduce artificial loss channels. These may e.g. deplete the probabilities $\tilde p_l$ in \cref{eq:reddenmat} beyond the true effect induced by the presence of the environment.

In general \cref{eq:Lindblad} can be expressed in the coordinate space basis of the Hilbert space, where the matrix elements of the reduced density matrix $\rho(   {\bm x}_1,{\bm x}_2, \ldots, {\bm y}_2,{\bm y}_1,t )=\langle {\bm x}_1,{\bm x}_2, \ldots|  \hat\rho_{\rm S}(t) |  \ldots , {\bm y}_2,{\bm y}_1\rangle\in\mathbb{C}$ form a complex function with a dependence on twice the number of coordinates ${\bm x}_i,{\bm y}_i\in\mathbb{R}^3$ as are particles present in the system. The ensuing partial differential equation remains linear in the density matrix, but contains both spatially varying and complex valued coefficients and derivative terms for each coordinate
\begin{align}
&i\frac{d}{dt} \langle {\bm x}_1,{\bm x}_2, \ldots|  \hat\rho_{\rm S}(t) |  \ldots , {\bm y}_2,{\bm y}_1\rangle  = i\frac{\partial}{\partial t}\rho(   {\bm x}_1,{\bm x}_2, \ldots, {\bm y}_2,{\bm y}_1,t ) \label{eq:genLinbladcoord} \\
\nonumber &= F \Big[  {\bm x}_1,{\bm x}_2, \ldots, {\bm y}_2,{\bm y}_1,  {\bm \nabla}_{x1},{\bm \nabla}_{x2}, \ldots, {\bm \nabla}_{y2},{\bm \nabla}_{y1},t\Big] \rho(   {\bm x}_1,{\bm x}_2, \ldots, {\bm y}_2,{\bm y}_1,t ).
\end{align}
The computational challenge, which we address in this paper, lies in discretizing \cref{eq:genLinbladcoord} and implementing it with a stable and accurate numerical procedure, which in addition guarantees that the properties in \cref{eq:Lindbprop} are preserved. Using an arbitrary complex test function $f({\bm x}_1, {\bm x}_2,\ldots) \in\mathbb{C}$ and denoting by $\delta^{(3)}({\bm x}_1-{\bm y}_1)$ the three-dimensional delta function, these properties can be formulated in terms of the matrix elements as follows.
\begin{align}
&{\rm Positivity:}\quad \forall f({\bm x}_1, {\bm x}_2,\ldots) \in\mathbb{C}, \\
\nonumber&\int d^3x_1 d^3 x_2 \ldots d^3 y_2 d^3y_1 f({\bm x}_1, {\bm x}_2,\ldots)^* \rho({\bm x}_1, {\bm x}_2, \ldots, {\bm y}_2, {\bm y}_1, t)f({\bm y}_1, {\bm y}_2,\ldots)\geq 0\\
\nonumber \\
&{\rm Hermiticity:}\quad \rho({\bm y}_1, {\bm y}_2, \ldots, {\bm x}_2, {\bm x}_1, t)^*=\rho({\bm x}_1, {\bm x}_2, \ldots, {\bm y}_2, {\bm y}_1, t)\\
\nonumber\\
&{\rm Unit\ trace:}\\
\nonumber&\int d^3x_1 d^3 x_2 \ldots d^3 y_2 d^3y_1 \delta^{(3)}({\bm x}_1-{\bm y}_1)\ldots \rho({\bm x}_1, {\bm x}_2, \ldots, {\bm y}_2, {\bm y}_1, t)=1
\end{align}

The numerical treatment of initial-boundary value problems (IBVPs), among them the Navier-Stokes and Schr\"odinger-like equations, such as \cref{eq:genLinbladcoord}, has seen significant progress over the past decade with the development and refinement of summation-by-parts (SBP) difference operators (for reviews see e.g. \cite{svard2014review,fernandez2014review,lundquist2014sbp}). As these operators build upon the finite difference approach (although they can be formulated for many other schemes, see \cite{nordstrom2012weak,nordstrom2003finite,carpenter2014entropy,carpenter1996spectral,castonguay2013energy,huynh2007flux,ranocha2016summation,gassner2013skew,hesthaven1996stable}) they are straight forward to implement and their numerical evaluation cost is low. The fact that they mimic the integration by parts property of the continuum theory facilitates proofs of stability, e.g. when deploying SBP operators in time stepping approaches for computational fluid dynamics \cite{nordstrom2017roadmap}. After the development of SBP operators for first derivatives, higher derivative approximations \cite{mattsson2004summation,mattsson2014diagonal} have been derived. More recently the SBP technique has also been applied to derivatives in time direction \cite{lundquist2014sbp,nordstrom2013summation,nordstrom2016summation}. While in this study only periodic boundary conditions will be deployed, the SBP operators can easily accommodate non-trivial boundary conditions (in the weak sense) via the Simultaneous Approximation Term (SAT) technique \cite{carpenter1994time}.

To make the paper self-contained, we provide a brief introduction to SBP operators and recommend \cite{svard2014review,fernandez2014review} for extensive reviews. Let the domain $[x_L,x_R]$ be discretized with $N+1$ equidistant grid points $x_i =x_L + i \Delta x $, $i = 0,\dots, N$, where $\Delta x= (x_R- x_L)/N$. Denote by $\mathfrak{u}(t) = [u_0, \dots, u_N]^\top$ the vector containing the function $u(t,x)$ evaluated at spatial grid points at time $t$.  The approximation of the spatial derivative is given by
\begin{equation*}
   D \mathfrak{u} \approx
   \mathfrak{u}_x
   \, ,
\end{equation*}
where $\mathfrak{u}_x$ contains the analytical derivative evaluated on the grid. For two functions $\mathfrak{u},\mathfrak{v}$ defined on the grid, we have 
\begin{equation*}
  (\mathfrak{u},\mathfrak{v})_H = 
  \mathfrak{u}^\top H \mathfrak{v},
  \quad 
  \|\mathfrak{u}\|^2_H = (\mathfrak{u},\mathfrak{u})_H 
  \, ,
\end{equation*}
where the matrix $H$ is diagonal, positive definite and defines an inner product and a corresponding norm.
Furthermore, the differentiation operator $D$ satisfies the SBP property
\begin{equation}
  (\mathfrak{v}, D \mathfrak{u})_H = 
  -(\mathfrak{u}, D \mathfrak{v})_H + 
  \mathfrak{u}^\top(E_N - E_0) \mathfrak{v}
  \, ,
  \label{eq:SBPprop}
\end{equation}
where $E_N={\rm diag}[0,\ldots,1]$ and $E_0={\rm diag}[1,\ldots,0]$.

In the second order case, $H$ is the composite trapezoidal rule and $D$ is the standard stencil for the symmetric central difference in the interior and appropriate forward and backward stencils at the boundaries: 
\begin{equation}
\nonumber H=\Delta x \left[ \begin{array}{ccccc} 1/2 & & & & \\ &1 & & &\\ & &\ddots && \\ &&&1&\\ &&&&1/2 \end{array} \right],
\quad 
D=
\frac{1}{2 \Delta x}
\left[ \begin{array}{ccccc} -2 &2 & & &\\ -1& 0& 1& &\\ & &\ddots && \\ &&-1&0&1\\ &&&-2&2 \end{array} \right].
\end{equation}
In the periodic case, the operators simplify to
\begin{equation*}
H= \Delta x 
\left[ \begin{array}{ccccc} 1 & & & & \\ &1 & & &\\ & &\ddots && \\ &&&1&\\ &&&& 1 \end{array} \right]
,
\quad 
D=
\frac{1}{2 \Delta x}
\left[ \begin{array}{ccccc} 0 &1 & & & -1\\ -1& 0& 1& &\\ & &\ddots && \\ &&-1&0&1\\ 1 &&&-1&0 \end{array} \right].
\end{equation*}

The SBP property already provides a crucial ingredient in the formulation of stable approximations of IBVPs, such as \cref{eq:genLinbladcoord}. In order to also preserve the trace of a relevant class of Lindblad equations, we will show that another continuum property of derivatives needs to be fulfilled: reparametrization neutrality.

Reparametrization neutrality refers to the fact that among a set of derivatives with respect to different variables, we may freely change to derivatives expressed in linear combinations of these variables. As a concrete example take $x$ and $y$ and the corresponding derivatives $\frac{d}{dx}$ and $\frac{d}{dy}$. Considering instead $z=x-y$ and $z^\prime=x+y$ we may reexpress \[ \frac{d}{dx} = \Big( \frac{d}{dz^\prime} +\frac{d}{dz} \Big) \text{ and } \frac{d}{dy} = \Big( \frac{d}{dz^\prime} -\frac{d}{dz} \Big). \] For the derivation of the trace conservation in the Lindblad equation we will need to switch from expressions in $x$ and $y$ to expressions in $z$ and $z^\prime$, which, when discretized with the conventional symmetric finite difference operator, turns out to be impossible. Hence we set out to define a novel finite difference operator, which besides the summation-by-parts property also remains neutral under reparametrization.

We proceed in \cref{sec:QQbarLindblad} by formulating an explicit expression for \cref{eq:genLinbladcoord} for the dissipative dynamics of a heavy quarkonium particle, interacting with a hot environment. We will discuss the preservation of the defining properties of the density matrix in the continuum and pinpoint where it fails after discretization. Using this insight we will in \cref{sec:RNSBPOp} construct a novel reparametrization neutral SBP operator, which, as we show, retains the continuum properties in the discretized evolution equations. In \cref{sec:NumRes} we will present numerical results from the simulation of quarkonium dissipative dynamics, showcasing the successful preservation of the continuum properties of the density matrix. We close with a brief conclusion and outlook in \cref{sec:conclout}

\section{Quarkonium Lindblad master equation}
\label{sec:QQbarLindblad}

As a concrete example of a Lindblad equation describing dissipative dynamics of a phenomenologically relevant system, we present the case of heavy quark--anti-quark bound states, so called heavy quarkonium at high temperature. The dissipative dynamics of these bound states immersed in a thermal medium play an important role in our understanding of heavy-ion collisions carried out e.g. at the Large Hadron Colldier at the CERN Laboratory. In such collisions, nuclei of heavy atoms are smashed into each other at ultra-relativistic momenta, so that the protons and neutrons making up the nuclei become compressed and heated to temperatures beyond $200,000\times$ the temperatures present in the core of the sun.  In turn they melt into their microscopic constituents, the light quarks and gluons, which form a so called quark-gluon-plasma (QGP). Some of the kinetic energy in the original projectiles is converted into additional particles, such as heavy quark--anti-quark pairs, which may form quarkonium bound states that find themselves immersed in the approximately locally thermal QGP. Using the quantum field theory quantum-chromo-dynamics (QCD) one wishes to understand how the interaction between the quarkonium and the medium affects their binding properties. If we translate these in-medium modifications into changes in the quarkonium particle production rate, we may use their measured yields in heavy-ion collisions to deduce the properties of the QGP created therein.

In \cite{Akamatsu:2012vt,Akamatsu:2014qsa} the Lindblad equation for a single quarkonium particle at high temperature was derived. Here the term high refers to the fact that the strong interactions between quarks and gluons become weaker at high energy scales and thus a weak-coupling expansion could be deployed. The explicit expressions for the subsystem Hamiltonian and the Lindblad operators in terms of the three-dimensional coordinates of the quark and anti-quark ${\bm x}_Q$ and ${\bm x}_{\bar{Q}}$ and their momenta ${\bm p}_Q$,${\bm p}_{\bar Q}$ read
\begin{subequations}
\begin{align}
\tilde H_S &= \frac{\hat {\bm p}_Q^2+\hat {\bm p}_{\bar Q}^2}{2M} +
\left[V(\hat {\bm x}_Q-\hat {\bm x}_{\bar Q})-\frac{1}{8MT}\left\{
(\hat{\bm p}_Q-\hat{\bm p}_{\bar Q}), {\bm \nabla} D(\hat{\bm x}_Q-\hat{\bm x}_{\bar Q})
\right\}\right](\ta \otimes \tac),\\
\hat L_{{\bm k},a}&=\sqrt{\frac{\tilde D({\bm k})}{2L^3}}
\left[
e^{\frac{i{\bm k}\cdot\hat {\bm x}_Q}{2}}
\left(1-\frac{{\bm k}\cdot\hat {\bm p}_Q}{4MT}\right)
e^{\frac{i{\bm k}\cdot\hat {\bm x}_Q}{2}}(\ta\otimes \1)
-e^{\frac{i{\bm k}\cdot \hat {\bm x}_{\bar Q}}{2}}
\left(1-\frac{{\bm k}\cdot\hat {\bm p}_{\bar Q}}{4MT}\right)
e^{\frac{i{\bm k}\cdot \hat {\bm x}_{\bar Q}}{2}}(\1\otimes\tac)
\right].
\end{align}
\label{eq:Lindblad2}
\end{subequations}
The Lindblad operators in this case depend on a continuous momentum variable ${\bm k}$ and a discrete "color" index $a$ related to the triple valued charge of strongly interacting particles. I.e. each "entry" of the Hamiltonian and the Lindblad operators corresponds to a $6\times 6$ matrix built from direct products of the $3\times 3$ Gell-Mann matrices denoted by $t^a$.

The form of $\hat L_{{\bm k},a}$ is intuitively understandable: there are two terms, one for the interaction of the medium with the quark-, another term for the interaction with the anti-quark constituent making up the quarkonium particle. The relative minus sign between the two terms in the second line of \cref{eq:Lindblad2} indicates that we are dealing with a system consisting of a particle and anti-particle. All aspects of the dynamics are captured in two quantities, the real-valued potential $V({\bm x})$ and the dissipation kernel $D({\bm x})$. These two quantities are intimately related to the real- and imaginary part of the proper real-time heavy-quark potential computed perturbatively in \cite{Laine:2006ns,Beraudo:2007ky,Brambilla:2008cx} and in numerical simulations of the strong interactions (so called lattice QCD) in \cite{Rothkopf:2011db,Burnier:2014ssa,Burnier:2015tda,Petreczky:2018xuh}. It has been shown that the effects of dissipation are encoded in the momentum ${\bm k}$ dependent terms in the parentheses, without which the dissipationless evolution of the recoilless-limit is obtained.

The description of the two-body quarkonium system may be simplified by going over to relative and center of mass coordinates for the quark--anti-quark degrees of freedom
\begin{align}
\hat {\bm x} = \frac{\hat{\bm x}_Q+\hat{\bm x}_{\bar Q}}{2}, \quad
\hat {\bm y} = \hat{\bm x}_Q-\hat{\bm x}_{\bar Q}, \quad
\hat {\bm P} = \hat{\bm p}_Q+\hat{\bm p}_{\bar Q}, \quad
\hat {\bm p} = \frac{\hat{\bm p}_Q-\hat{\bm p}_{\bar Q}}{2}.
\end{align}
Tracing out also the center of mass coordinate, it was shown in \cite{Miura:2019ssi} that the following relative coordinate operators ensue
\begin{align}
\tilde H_S^{\rm rel} &= 
\frac{\hat {\bm p}^2 }{M} + V(\hat {\bm x}) (\ta \otimes \tac) - \frac{1}{4MT} \left\{
\hat{\bm p},  {\bm \nabla}D(\hat{\bm x})\right\},\\
\nonumber\hat L_{{\bm k},a}^{\rm rel}&=\sqrt{\frac{\tilde D({\bm k})}{2L^3}}
\left[1-\frac{{\bm k}}{4MT}\cdot \left( \frac{1}{2} {\bm P} +\hat{{\bm p}}\right)\right] e^{\frac{i{\bm k}{\bm r}}{2}}(\ta\otimes \1)\\
&-\sqrt{\frac{\tilde D({\bm k})}{2L^3}}\left[1-\frac{{\bm k}}{4MT}\cdot \left( \frac{1}{2} {\bm P} -\hat{{\bm p}}\right)\right] e^{-\frac{i{\bm k}{\bm r}}{2}}(\1\otimes\tac).
\end{align}
\label{eq:Lindblad3}
In the following we will further simplify the description by considering the quarkonium particle to be at rest ${\bm P}=0$ and neglecting the explicit matrix structure of the Lindblad operators and the Hamiltonian, essentially setting $t^a$ to unity and retaining only a single color component.

Evaluating the ensuing Lindblad equation in the coordinate space basis for the relative coordinates, we obtain the following partial differential equation with spatially and potentially temporally varying coefficient terms 
\begin{align}
\partial_t \rho^{\rm rel}({\bm x},{\bm y},t)=i \Big[ \frac{{\bm \nabla}_x^2}{M} -V({\bm x}) \Big]\rho^{\rm rel}({\bm x},{\bm y},t)  - i \Big[ \frac{{\bm \nabla}_y^2}{M} - V({\bm y})\Big]&\rho^{\rm rel}({\bm x},{\bm y},t)\label{eq:mastereq3d}\\
\nonumber+\Big[ 2F_1\Big( \frac{ {\bm x}-{\bm y} }{2} \Big) - 2F_1\big(  {\bm 0} \big) + F_1\big(  {\bm x} \big) + F_1\big(  {\bm y} \big) - 2F_1\Big( \frac{ {\bm x}+{\bm y} }{2} \Big) \Big] &\rho^{\rm rel}({\bm x},{\bm y},t)\\
\nonumber-\Big[  \frac{ ({\bm \nabla}_x^2)^2 A({\bm x}) }{4M^2}  +  \frac{ ({\bm \nabla}_y^2)^2 A({\bm y}) }{4M^2}  \Big] &\rho^{\rm rel}({\bm x},{\bm y},t)\\
\nonumber+\Big[ 2{\bm F}_2\Big( \frac{ {\bm x}-{\bm y} }{2} \Big) + 2{\bm F}_2\big(  {\bm x} \big) - 2{\bm F}_2\Big( \frac{ {\bm x}+{\bm y} }{2} \Big)  -  {\bm \nabla}_x\frac{ ({\bm \nabla}_x^2) A({\bm x}) }{M^2} \Big] &{\bm \nabla}_x\rho^{\rm rel}({\bm x},{\bm y},t)\\
\nonumber+\Big[ -2{\bm F}_2\Big( \frac{ {\bm x}-{\bm y} }{2} \Big) + 2{\bm F}_2\big(  {\bm y} \big) - 2{\bm F}_2\Big( \frac{ {\bm x}+{\bm y} }{2} \Big) -  {\bm \nabla}_y\frac{ ({\bm \nabla}_y^2) A({\bm y}) }{M^2} \Big] &{\bm \nabla}_y\rho^{\rm rel}({\bm x},{\bm y},t)\\
\nonumber+\Big[ 2F_3^{ij}\Big( \frac{ {\bm x}-{\bm y} }{2} \Big) + 2F_3^{ij}\Big( \frac{ {\bm x}+{\bm y} }{2} \Big)\Big]&\nabla_x^i\nabla_y^j \rho^{\rm rel}({\bm x},{\bm y},t)\\
\nonumber+\Big[ \frac{1}{3} F^{kk}_3(  {\bm 0} \big) \delta^{ij} + F^{ij}_3(  {\bm x} \big) \Big]& \nabla_x^i\nabla_x^j \rho^{\rm rel}({\bm x},{\bm y},t)\\
\nonumber+\Big[ \frac{1}{3} F^{kk}_3(  {\bm 0} \big) \delta^{ij}  + F^{ij}_3(  {\bm y} \big) \Big]& \nabla_y^i\nabla_y^j \rho^{\rm rel}({\bm x},{\bm y},t).
\end{align}
The vectorial derivatives acting on the coordinate vectors ${\bm x}$ and ${\bm y}$ are  defined as ${\bm \nabla}_x=(\frac{\partial}{\partial x_1},\frac{\partial}{\partial x_2},\frac{\partial }{\partial x_3})$ and ${\bm \nabla}_y=(\frac{\partial }{\partial y_1},\frac{\partial }{\partial y_2},\frac{\partial }{\partial y_3})$ respectively and their components denoted by $\nabla_x^i$ and $\nabla_y^i$.

The scalar function $F_1$, the vectorial ${\bm F}_2$ and the tensorial $F^{ij}_3$ have been introduced to conveniently summarize the contributions arising from the dissipation kernel
\begin{align}
F_1\big(  {\bm x} \big) &= \Big[ D( {\bm x} ) + \frac{ {\bm \nabla}^2_x D( {\bm x} ) }{4MT}+ \frac{ \big({\bm \nabla}^2_x\big)^2 A( {\bm x} ) }{8M^2}\Big],\\
{\bm F}_2\big(  {\bm x} \big) &= {\bm \nabla}_x \Big[  \frac{ D( {\bm x} ) }{4MT}+ \frac{ {\bm \nabla}^2_x A( {\bm x} ) }{4M^2}\Big],\\
F^{ij}_3\big(  {\bm x} \big) &=-\nabla_x^i\nabla_x^j\Big[ \frac{ A( {\bm x} ) }{2M^2}\Big] \label{eq:defF3}
\end{align}
and the function $A({\bm x})=D({\bm x})/8T^2$. Up to this point no approximation beyond the weak coupling expansion and time coarse graining enter our description.

While our long-term goal is to solve the full three-dimensional dynamics of quarkonium based on \cref{eq:mastereq3d}, we restrict ourselves in this paper to the one-dimensional case. It already presents us with many of the relevant technical challenges, while requiring significantly less computational resources for its implementation. It furthermore simplifies the presentation, without compromising with the fundamental computational development.

In one dimension, derivatives in the $x$ and $y$ coordinate do not carry indices anymore and \cref{eq:mastereq3d} reduces to 
\begin{align}
\partial_t \rho^{\rm rel}(x,y,t)=i \Big[ \frac{1}{M}\frac{\partial^2}{\partial x^2} -V(x) \Big]\rho^{\rm rel}(x,y,t)  - i \Big[ \frac{1}{M}\frac{\partial^2}{\partial y^2} - V(y)\Big]&\rho^{\rm rel}(x,y,t)\label{eq:mastereq}\\
\nonumber+\Big[ 2F_1\Big( \frac{ x-y }{2} \Big) - 2F_1\big(  0 \big) + F_1\big( x \big) + F_1\big(  y \big) - 2F_1\Big( \frac{ x+y }{2} \Big) \Big] &\rho^{\rm rel}(x,y,t)\\
\nonumber-\Big[ \frac{\partial^2}{\partial x^2}  \frac{ \frac{\partial^2}{\partial x^2} A(x) }{4M^2}  +  \frac{\partial^2}{\partial y^2}\frac{ \frac{\partial^2}{\partial y^2} A(y) }{4M^2}  \Big] &\rho^{\rm rel}(x,y,t)\\
\nonumber+\Big[ 2F_2\Big( \frac{ x-y }{2} \Big) + 2F_2\big(  x \big) - 2F_2\Big( \frac{ x+y }{2} \Big)  -  \frac{\partial}{\partial x} \frac{ \frac{\partial^2}{\partial x^2} A(x) }{M^2} \Big] \frac{\partial}{\partial x}&\rho^{\rm rel}(x,y,t)\\
\nonumber+\Big[ -2F_2\Big( \frac{ x-y }{2} \Big) + 2F_2\big(  x \big) - 2F_2\Big( \frac{ x+y }{2} \Big) -  \frac{\partial}{\partial y} \frac{ \frac{\partial^2}{\partial y^2} A(y) }{M^2} \Big] \frac{\partial}{\partial y}&\rho^{\rm rel}(x,y,t)\\
\nonumber+\Big[ 2F_3\Big( \frac{ x-y }{2} \Big) + 2F_3\Big( \frac{ x+y }{2} \Big)\Big] \frac{\partial}{\partial x}\frac{\partial}{\partial y} & \rho^{\rm rel}(x,y,t)\\
\nonumber+\Big[  F_3(  0 \big)  + F_3(  x \big) \Big] \frac{\partial^2}{\partial x^2} \rho^{\rm rel}(x,y,t)+\Big[ F_3( 0 \big) + F_3(  y \big) \Big]\frac{\partial^2}{\partial y^2} &\rho^{\rm rel}(x,y,t).
\end{align}

Here the physics of the dissipation kernel $D(x)$ enters via three real-valued scalar functions, which, keeping in mind that $A(x)=D(x)/8T^2$, read
\begin{align}
F_1\big(  x\big) &= \Big[ D( x) + \frac{1 }{4MT} \frac{\partial^2}{\partial x^2}D(x) + \frac{ 1 }{8M^2}\frac{\partial^4}{\partial x^4} A(x ) \Big],\\
\nonumber F_2\big( x \big) &= \frac{ 1 }{4MT}\frac{\partial}{\partial x}D(x) + \frac{ 1 }{4M^2}\frac{\partial^3}{\partial x^3} A(x),\quad  F_3\big(  x \big) =- \frac{1}{2M^2} \frac{\partial^2}{\partial x^2} A(x).
\end{align}

\subsection*{Conservation of defining properties in one-dimension}

In order for positivity and hermiticity of the density matrix to be conserved in the Lindblad formalism, the function $D(x)$ in momentum space needs to be positive and real. As the dissipation kernel is supplied as external input, this property can be explicitly checked for and we will make sure it is fulfilled in the simulations that follow.  Here we focus on the preservation of the trace in \cref{eq:mastereq}, which presents the central challenge in its discretization. In the functional language, the trace over quantum states translates into an integration of $\rho^{\rm rel}(x, y, t)$  over $x$ and $y$ in the presence of a delta function $\delta(x-y)$.

Some of the terms in the trace over the first, second, fourth and fifth line of \cref{eq:mastereq} vanish identically, due to the arguments of the $V$ and $F$ functions being evaluated at $x=y$, e.g.
\begin{align}
T_1=\int dx \int dy  \, \delta(x-y) \Big[ &iV(x)-iV(y) +2F_1\Big( \frac{x-y}{2} \Big) - 2F_1\big(  0 \big) \\
\nonumber& + F_1\big(  x \big) + F_1\big(  y \big) - 2F_1\Big( \frac{ x+y }{2} \Big) \Big]\rho^{\rm rel}(x,y,t)=0,\\
T_2=\int dx \int dy  \, \delta(x-y) \Big[ & -2F_2\Big( \frac{ x+y }{2} \Big) \big( \frac{\partial}{\partial x} +\frac{\partial}{\partial y} \big)+ 2F_2\big( x\big)\frac{\partial}{\partial x}  \\
\nonumber & \quad\quad \quad \quad +2F_2\big( y \big)\frac{\partial}{\partial y} \Big]\rho^{\rm rel}(x,y,t)=0.
\end{align}
In addition, the following term from lines four and five vanishes 
\begin{align}
T_3=\int dx \int dy  \delta(x-y) \Big[ &2F_2\Big( \frac{x-y }{2} \Big) \big( \frac{\partial}{\partial x} -\frac{\partial}{\partial y}\big) \Big] \rho^{\rm rel}(x,y,t)=0.
\end{align}
This can be seen by inspecting the properties of the function $D(x)$ and the definition of $F_2$, in which the first and third derivative of $D(x)$ enters (remember $A(x)=D(x)/8T^2$). The derivation of the Lindblad equation from QCD in \cite{Akamatsu:2012vt,Akamatsu:2014qsa} leads to a function $D(x)$ that possesses a maximum around the origin and which is furthermore symmetric around the origin. Thus, when we take the first and third derivative of $D(x)$ at the origin, both contributions vanish, i.e. $F_2(0)=0$. (We ensure that this property is respected in our numerical simulations, see \cref{eq:parmsim2}.)

For some terms in the trace over \cref{eq:mastereq} we need to apply integration by parts. Take e.g. the derivatives in the first line
\begin{align}
T_4=\int dx \int dy  \delta(x-y) \left[ i \frac{1}{M}\frac{\partial^2}{\partial x^2} - i \frac{1}{M}\frac{\partial^2}{\partial y^2} \right] &\rho^{\rm rel}(x,y,t)=0\label{eq:ibponlyterm}.
\end{align}
In order to show that these two contributions cancel each other, we need to be able to transform the double derivative in $x$ into a corresponding derivative in $y$, which is possible due to the delta-function. We get
\begin{align}
\nonumber &\int dx \int dy \delta(x-y)\frac{\partial^2}{\partial x^2} \rho^{\rm rel}(x,y,t)=\int dx \int dy \frac{\partial^2}{\partial x^2} \delta(x-y) \rho^{\rm rel}(x,y,t)
\end{align}
from twice integrating by parts and the fact that we consider periodic boundary conditions. Now we exploit the anti symmetry of the argument of the delta function to replace $\frac{\partial}{\partial x}$ by $-\frac{\partial}{\partial y}$ twice. (One may use a regularized form of the delta-function here to make the operation mathematically well defined.) Applying integration by parts twice again, we thus obtain
\begin{align}
\nonumber &\int dx \int dy \frac{\partial^2}{\partial y^2} \delta(x-y) \rho^{\rm rel}(x,y,t)=\int dx \int dy \delta(x-y)\frac{\partial^2}{\partial y^2} \rho^{\rm rel}(x,y,t),
\end{align}
which hence cancels the term in \cref{eq:ibponlyterm}.

Similarly, by application of integration by parts in the trace over terms in line six and seven of \cref{eq:mastereq}, we find that the following combination vanishes
\begin{align}
T_5=\int dx \int dy  \delta(x-y) \Big[ 2F_3\Big(\frac{x-y }{2} \Big)&\frac{\partial}{\partial x}\frac{\partial}{\partial y} \\
\nonumber +&F_3(0) \big(\frac{\partial^2}{\partial x^2}+\frac{\partial^2}{\partial y^2}\big) \Big] &\rho^{\rm rel}(x,y,t)=0.
\end{align}
The need to perform integration by parts in the continuum theory tells us that the difference operators, to be deployed for discretization of \cref{eq:mastereq}, need to fulfil the summation-by-parts property.

For the other terms to vanish, additional continuum properties of derivatives are required. Let us start with the remaining $F_3$ terms in the last two lines of \cref{eq:mastereq}
\begin{align}
T_6=\int dx \int dy  \delta(x-y) \Big[ \underbracket{2F_3\Big( \frac{ x+y }{2} \Big)\frac{\partial}{\partial x}\frac{\partial}{\partial y}}_{T_{61}}  &+   F_3(  x \big)  \frac{\partial^2}{\partial x^2} \label{eq:termC}\\
\nonumber +&  F_3(  y \big) \frac{\partial^2}{\partial y^2} \Big] \rho^{\rm rel}(x,y,t).
\end{align}
This expression will not vanish by itself but instead produces a remnant term, which in turn will cancel with the $A$ dependent contributions in the trace over \cref{eq:mastereq}. At first sight moving around the derivatives on the first $F_3$ term (denoted as $T_{61}$) appears to involve the application of the product rule which would lead to subsequent numerical complications, as splitting would be required \cite{nordstrom2006conservative}. Note however the particular form of this coefficient. It depends only on the sum $z^\prime=x+y$ of the coordinates, while the delta function only depends on $z=x-y$, which invites us to introduce
\begin{align}
\frac{\partial}{\partial x} & = \Big( \frac{\partial}{\partial z^\prime} + \frac{\partial}{\partial z} \Big),
& \frac{\partial}{\partial y} & = \Big(  \frac{\partial}{\partial z^\prime} - \frac{\partial}{\partial z} \Big), \label{eq:reparm}
\\
\frac{\partial}{\partial z^\prime} &= \frac{1}{2}\left( \frac{\partial}{\partial x} + \frac{\partial}{\partial y} \right), 
& \frac{\partial}{\partial z} & = \frac{1}{2}\left( \frac{\partial}{\partial x} - \frac{\partial}{\partial y} \right). \label{eq:reparm2}
\end{align}

Let us have a look at how the reparametrization property of the differentials of the two sets of coordinates in the continuum can be used to rewrite the mixed derivative term in \cref{eq:termC}. We obtain
\begin{align}
\nonumber 2 \frac{\partial}{\partial x} \frac{\partial}{\partial y} &= 2 \Big(\frac{\partial}{\partial z^\prime}\Big)^2 - 2 \Big(\frac{\partial}{\partial z}\Big)^2\\
\nonumber&= 2 \Big(\frac{\partial}{\partial z^\prime}\Big)^2 - \Big[ \Big(\frac{\partial}{\partial x}\Big)^2+\Big(\frac{\partial}{\partial y}\Big)^2 -2\Big(\frac{\partial}{\partial z^\prime}\Big)^2 \Big]\\
\nonumber&= 4 \Big(\frac{\partial}{\partial z^\prime}\Big)^2 - \Big(\frac{\partial}{\partial x}\Big)^2-\Big(\frac{\partial}{\partial y}\Big)^2\\
&=2 
 \Big( \frac{\partial}{\partial x} + \frac{\partial}{\partial y} \Big)
\frac{\partial}{\partial z^\prime}- 
\Big(\frac{\partial^2}{\partial x^2}\Big)^2-\Big(\frac{\partial}{\partial y}\Big)^2.\label{eq:manipreqforTermc}
\end{align}

The result of \cref{eq:manipreqforTermc} can be directly applied to the mixed derivative term $T_{61}$ in \cref{eq:termC}:
\begin{equation}
T_{61} =\int dx \int dy  \delta(z) F_3\Big( \frac{z'}{2} \Big)\left( 2 \Big(\frac{\partial}{\partial x} + \frac{\partial}{\partial y} \Big)\frac{\partial}{\partial z^\prime}  - \frac{\partial^2}{\partial x^2}-\frac{\partial^2}{\partial y^2}\right)\rho(x,y,t)
\, .
\label{eq:T61}
\end{equation}
Since $z'/2 = x = y$ along the trace, inserting Eq. \eqref{eq:T61} into Eq. \eqref{eq:termC} yields 
\begin{equation}
\begin{aligned}
T_6 & = 
\int dx \int dy \delta(z) F_3 \Big(\frac{z'}{2}\Big) 
 2\Big( \frac{\partial}{\partial x} + \frac{\partial}{\partial y} \Big) \frac{\partial}{\partial z^\prime}\rho(x,y,t)
 \\
& = 
-\int dx \int dy \Big( \frac{\partial}{\partial x} + \frac{\partial}{\partial y} \Big)\delta(z) F_3 \Big(\frac{z'}{2}\Big) 
 2 \frac{\partial}{\partial z^\prime}\rho(x,y,t)
\\
& = 
-\int dx \int dy \delta(z) 2\frac{\partial}{\partial z^\prime} F_3 \Big(\frac{z'}{2}\Big) 
2 \frac{\partial}{\partial z^\prime}\rho(x,y,t)^{\rm rel}
\\
&= - \int dx \int dy  \delta(z) 2 
\left[\frac{\partial}{\partial z^\prime} F_3\big(  \frac{z^\prime}{2} \big) \frac{\partial}{\partial x} + \frac{\partial}{\partial z^\prime} F_3\big( \frac{z^\prime}{2}\big)\frac{\partial}{\partial y} \right]\rho^{\rm rel}(x,y,t).
\label{eq:termCmanip}
\end{aligned}
\end{equation}
In Eq. \eqref{eq:termCmanip}, we carried out one integration by parts in both $x$ and $y$. Since $\delta(x-y)$ only depends on $z$ and not $z^\prime$, the derivative $\partial/\partial z^{\prime}$ after integration by parts acts solely on the $F_3$ term. 

Let us suggestively rewrite the derivatives over $z^\prime$ in terms of a variable $\xi$, which, due to the presence of the delta function, we will later identify with $x$ or $y$
\begin{align}
T_{6}= - \int dx \int dy  \delta(x-y) \Big[ \frac{\partial}{\partial \xi} & \left.F_3\big( \xi \big)\right|_{\xi=\frac{x+y}{2}} \frac{\partial}{\partial x} \\
\nonumber +& \left. \frac{\partial}{\partial \xi} F_3\big( \xi \big)\right|_{\xi=\frac{x+y}{2}}\frac{\partial}{\partial y} \Big] \rho^{\rm rel}(x,y,t).\label{eq:xideriv}
\end{align}
Note that the factor 2 from \cref{eq:termCmanip} has disappeared due to the application of the chain rule. Using the definition in \cref{eq:defF3}, we can turn these two $F_3$ terms into derivatives acting on the $A$ function 
\begin{align}
T_{6}= \int dx \int dy  \delta(x-y) \frac{1}{2M^2}\Big[  \frac{\partial^3}{\partial \xi^3}& \left.A\big( \xi \big)\right|_{\xi=\frac{x+y}{2}} \frac{\partial}{\partial x} \\
\nonumber+& \left. \frac{\partial^3}{\partial \xi^3} A\big( \xi \big)\right|_{\xi=\frac{x+y}{2}}\frac{\partial}{\partial y} \Big] \rho^{\rm rel}(x,y,t).\label{eq:termCfurther}
\end{align}
Comparing the above expression to the fourth and fifth line of \cref{eq:mastereq} we see that it has a form very similar to the A terms present there
\begin{align}
T_7=\int dx \int dy  \delta(x-y) \Big[  -  \Big(\frac{\partial}{\partial x} \frac{ \frac{\partial^2}{\partial x^2} A(x) }{M^2}\Big) & \frac{\partial}{\partial x}\\
\nonumber-& \Big(\frac{\partial}{\partial y} \frac{ \frac{\partial^2}{\partial y^2} A(y) }{M^2}\Big)\frac{\partial}{\partial y} \Big] &\rho^{\rm rel}(x,y,t).
\end{align}
Making use of the fact that in \cref{eq:termCfurther}, inside the trace, we can replace $\frac{\partial^3}{\partial \xi^3}$ by either $\frac{\partial^3}{\partial x^3}$ or $\frac{\partial^3}{\partial y^3}$, we can cancel parts of the terms in $T_7$.
Note that due to the factor $1/2$ present in \cref{eq:xideriv} we obtain a rest term when summing $T_{6}$ and $T_7$.

In order to bring this rest term into the form necessary to cancel the last remaining A terms in \cref{eq:mastereq} it may be conveniently expressed in the $\xi$ derivatives of \cref{eq:termCfurther}
\begin{align}
T_{6}+T_7= -\int dx \int dy  \delta(x-y) \frac{1}{2M^2}\Big[  & \left.\frac{\partial^3}{\partial \xi^3} A\big( \xi \big)\right|_{\xi=\frac{x+y}{2}} \frac{\partial}{\partial x}\\
\nonumber + & \left. \frac{\partial^3}{\partial \xi^3} A\big( \xi \big)\right|_{\xi=\frac{x+y}{2}}\frac{\partial}{\partial y} \Big] \rho^{\rm rel}(x,y,t).
\end{align}
Let us now summarize the two terms as a single expression with derivative in $z^\prime$
\begin{align}
T_{6}+T_7=& -\int dx \int dy  \delta(x-y) \frac{1}{M^2}\Big[  \left.\frac{\partial^3}{\partial \xi^3} A\big( \xi \big)\right|_{\xi=\frac{x+y}{2}} \frac{\partial}{\partial z^\prime}  \Big] \rho^{\rm rel}(x,y,t).
\end{align}
We may now carry out integration by parts in $z^\prime$ (by shifting $\frac{\partial}{\partial z'}$ to $\frac{1}{2}(\frac{\partial}{\partial x} + \frac{\partial}{\partial y})$), exploiting that $\delta(x-y)$ does not depend on $z^\prime$. Using again the fact that inside the trace we can replace $\frac{\partial^3}{\partial \xi^3}$ by either $\frac{\partial^3}{\partial x^3}$ or $\frac{\partial^3}{\partial y^3}$ we arrive at the final expression
\begin{align}
T_{6}+&T_7=\\
\nonumber=& \int dx \int dy  \delta(x-y) \frac{1}{M^2}\Big[  \Big(\frac{1}{2} \frac{\partial}{\partial x}+\frac{1}{2} \frac{\partial}{\partial y}\Big)\left.\frac{\partial^3}{\partial \xi^3} A\big( \xi \big)\right|_{\xi=\frac{x+y}{2}}   \Big] \rho^{\rm rel}(x,y,t)\\
\nonumber =&\int dx \int dy  \delta(x-y) \frac{1}{M^2}\Big[ \frac{1}{4}\frac{\partial^4}{\partial x^4}A(x) + \frac{1}{4}\frac{\partial^4}{\partial y^4}A(y)\Big] \rho^{\rm rel}(x,y,t).
\end{align}
which cancels identically with the only remaining A terms in the third line of \cref{eq:mastereq}. This concludes the explicit demonstration that the one-dimensional Lindblad master equation preserves the unit trace property of the reduced density matrix.
\begin{remark}
The lesson learned for the discretization of \cref{eq:mastereq} is that the application of the product rule is not necessary in order to conserve the trace, as long as the reparametrization property of \cref{eq:reparm} is fulfilled. This bodes well, as it is well known that discretizations of the difference operator in general violate the product rule, even if they fulfill e.g. the summation by parts property \cite{nordstrom2006conservative}. To summarize, we need difference operators that can integrate by parts in $x,y$ and differentiate in $z,z'$.
\end{remark}

However, the standard difference operator is not neutral under reparametrization, as is evidenced by (assuming the same equidistant discretization $\Delta$ in ${\bm x}$ and ${\bm y}$)
\begin{align}
&( \mathds{D}^{\rm naive}_x+ \mathds{D}^{\rm naive}_y) f(x,y) =\\
\nonumber&\frac{1}{2\Delta }\big( f(x +\Delta,y)-f(x -\Delta,y) + f(x, y+ \Delta )-f(x, y -\Delta)\big)\\
\nonumber&\neq \mathds{D}^{\rm naive}_{z^\prime} f(x,y) =  \frac{1}{2\Delta_z }\big( f(x +\Delta_z/2,y+\Delta_z/2)  - f(x -\Delta_z/2,y-\Delta_z/2)    \big).\label{eq:nonrepneutral}
\end{align}
Our goal thus is to construct a SBP operator that remains neutral under the reparametrization $({\bm x},{\bm y})\to ({\bm z},{\bm z}^\prime)$.

\section{A reparametrization-neutral summation by parts (RN-SBP) operator}
\label{sec:RNSBPOp}

We proceed to construct a novel SBP difference operator, which strictly implements the reparametrization property \cref{eq:reparm}, restricting ourselves here to the case of an equidistantly discretized function $\rho(x,y)$ with $\Delta x =\Delta y=\Delta$ and periodic boundary conditions. After choosing an explicit ordering of the discretized density matrix $\rho$ on the now two-dimensional (x,y) grid we introduce the shift operators $S_+$ and $S_-$ as follows
\begin{align}
\rho = \left[ \begin{array}{c} \rho(x_0,y_0) \\ \rho(x_0,y_1) \\ \vdots \\  \rho(x_0,y_{N-1}) \\ \rho(x_1,y_0)\\ \rho(x_1,y_1) \\ \vdots \\\rho(x_{N-1},y_{N-1}) \end{array}  \right], \quad S_+=\left[ \begin{array} {ccccc} 0 & 0 & 0& \hdots & 1 \\ 1 & 0 & 0& & 0 \\0 & 1 & 0& & 0 \\0 & 0 & 1& & 1 \\ \vdots & \vdots & 0& \ddots & \vdots \\ 0 & 0 & \hdots& 1 & 0 \end{array}\right], S_-=S_+^T.
\end{align}
Note that the consecutive application of the two shift operators yields $S_+ S_- = S_- S_+ = \mathbb{1}$, where $\mathbb{1}$ is the identity. Our strategy is to combine these shifts together with regular SBP difference operators to achieve the desired reparametrization neutrality. Indeed, $I_x\otimes S_+$ shifts rows upward on the discretized grid, while $ S_+\otimes I_y$ shifts columns to the right, with the inverse operations naturally being  $I_x\otimes S_-$ and $ S_-\otimes I_y$ respectively.

For periodic boundary conditions the simplest second order periodic SBP operator is constructed using the integration prescription $H=\Delta \mathbb{1}$ and 
\begin{align}
Q=\left[ \begin{array}{ccccc} 0 &1 & & &-1\\ -1& 0& 1& &\\ & &\ddots && \\ &&-1&0&1\\1 &&&-1&0 \end{array} \right],
\end{align}
where $D\equiv H^{-1} Q$. Since $Q + Q^\top = 0$, we get
\begin{equation}
    \label{eq:periodic_sbp}
    \mathfrak{u}^\top HD \mathfrak{v}
        = \mathfrak{u}^\top Q \mathfrak{v} 
        = \mathfrak{u}^\top \left(Q + Q^\top - Q^\top\right) \mathfrak{v}
        = -\mathfrak{u}^\top Q^\top \mathfrak{v}
        = -\mathfrak{v}^\top D \mathfrak{u} ,
\end{equation}
and the SBP property simplifies to $(\mathfrak{u},D\mathfrak{v})_H=-(D\mathfrak{u},\mathfrak{v})_H$. 

As indicated by \cref{eq:nonrepneutral}, for the reparametrization property to hold we need to express the $x$- and $y$- derivative of the function $\rho_{i,j}$ in terms of its values at the neighboring diagonal corners, i.e. in the variables $x+y$ and $x-y$. This is possible if we compute the average of the naive finite differences once shifted up and down in rows and columns respectively. With this strategy we arrive at the definition of the following reparametrization neutral summation-by-parts operators (RN-SBP)
\begin{align}
\mathds{D}_x = \frac{1}{2} \Big( D\otimes S_+ +D \otimes S_-\Big),\quad  \mathds{D}_y = \frac{1}{2} \Big( S_+\otimes D  + S_- \otimes D \Big).
\end{align}
The explicit expressions applying $\mathds{D}_x $ and $\mathds{D}_y$ to the function $\rho$ at $(x_i,y_j)$ are
\begin{align}
&(\mathds{D}_x \rho)_{i,j} = \frac{1}{2} \Big( \frac{ \rho_{i+1,j+1}-\rho_{i-1,j+1} }{2\Delta} +  \frac{ \rho_{i+1,j-1}-\rho_{i-1,j-1} }{2\Delta} \Big), \label{eq:Dx_idx}\\
&(\mathds{D}_y \rho)_{i,j} = \frac{1}{2} \Big( \frac{ \rho_{i+1,j+1}-\rho_{i+1,j-1} }{2\Delta} +  \frac{ \rho_{i-1,j+1}-\rho_{i-1,j-1} }{2\Delta} \Big). \label{eq:Dy_idx}
\end{align}
This construction also preserves the summation by parts property, as we can write
\begin{align}
\mathds{D}_x=\mathds{H}^{-1}\mathds{Q}_x, \quad \mathds{D}_y=\mathds{H}^{-1}\mathds{Q}_y,
\end{align} 
with $\mathds{H} = H\otimes H$, and
\begin{align}
&\mathds{Q}_x=\frac{1}{2}\big(Q\otimes HS_+ + Q\otimes H S_-\big), \quad \mathds{Q}_y=\frac{1}{2}\big(HS_+ \otimes Q + H S_- \otimes Q \big).
\end{align}
It follows that 
\begin{align}
\mathds{Q}_x+\mathds{Q}_x^T= \mathds{Q}_y+\mathds{Q}_y^T=0.
\end{align}
Hence, if $\mathfrak{u},\mathfrak{v}\in\mathds{R}^{N^2}$, then $(\mathfrak{u},\mathds{D}_x\mathfrak{v})_\mathds{H}=-(\mathds{D}_x\mathfrak{u},\mathfrak{v})_\mathds{H}$  (using the same argument as in \eqref{eq:periodic_sbp}). Similarly we get $(\mathfrak{u},\mathds{D}_y\mathfrak{v})_\mathds{H}=-(\mathds{D}_y\mathfrak{u},\mathfrak{v})_\mathds{H}$.

Let us next define the corresponding RN-SBP operators in $z$ and $z^\prime$ as in Eq. \eqref{eq:reparm2},
\begin{align}
\mathds{D}_{z^\prime}=\frac{1}{2} \Big( \mathds{D}_x+\mathds{D}_y \Big), \quad \mathds{D}_{z}=\frac{1}{2} \Big( \mathds{D}_x-\mathds{D}_y \Big),
\end{align}
which by \eqref{eq:Dx_idx}, \eqref{eq:Dy_idx} on the index level becomes
\begin{align}
&(\mathds{D}_{z^\prime} \rho)_{i,j}= \frac{ \rho_{i+1,j+1}-\rho_{i-1,j-1} }{2\Delta}, \quad (\mathds{D}_z \rho)_{i,j} = \frac{ \rho_{i-1,j+1}-\rho_{i+1,j-1} }{2\Delta}.
\end{align}
Naturally these new operators show the following behavior: if we have two functions $\mathfrak{u},\mathfrak{v}\in\mathds{R}^{N^2}$ and $\mathfrak{v}$ depends only on $z=(x-y)$, i.e. $\mathfrak{v}$ is constant along all lines $y=x+c$ then
\begin{align}
\mathds{D}_{z^\prime} (\mathfrak{u}\circ \mathfrak{v})=\mathfrak{v} \circ \mathds{D}_{z^\prime} \mathfrak{u},\label{eq:indepofzprime}
\end{align}
where $\circ$ denotes the Hadamard (elementwise) product. Similarly if $\mathfrak{u}$ depends only on $z^\prime=(x+y)$ we have
\begin{align}
\mathds{D}_{z} (\mathfrak{u} \circ \mathfrak{v})= \mathfrak{u} \circ \mathds{D}_{z} \mathfrak{v}.
\end{align}

Let us now show that the novel RN-SBP operators presented above allow us to implement in a discrete fashion, all manipulations deployed in the proof of the preservation of unit trace in \cref{sec:QQbarLindblad}.

The first set of manipulation we need to realize discretely is related to the terms $T_4,T_5$ and $T_6$. There, one must first perform summation by parts, which our novel RN-SBP operator implements by construction. Next we need to change x into y derivatives and vice versa via the delta function. Writing explicitly we get
\begin{align}
\mathds{D}_{x} \delta(z) &= \Big(\mathds{D}_{z^\prime}+\mathds{D}_{z}\Big)\delta(z)=\Big(-\mathds{D}_{z^\prime}+\mathds{D}_{z}\Big)\delta(z)=-\mathds{D}_{y}\delta(z).
\end{align}

Another type of operation is required to treat the $T_6$ term. In the continuum it amounts to the steps in \cref{eq:manipreqforTermc} which are identical for the RN-SBP operator
\begin{align}
\nonumber2 \mathds{D}_{x} \mathds{D}_{y}  &= \Big[ 2  \mathds{D}_{z^\prime}^2 - 2 \mathds{D}_{z}^2\Big] \\
\nonumber&=\Big[  2 \mathds{D}_{z^\prime}^2 - \Big\{ \mathds{D}_{x}^2 +\mathds{D}_{y}^2 -2 \mathds{D}_{z^\prime}^2 \Big\} \Big] \\
\nonumber&=\Big[ 4 \mathds{D}_{z^\prime}^2 - \mathds{D}_{x}^2 -\mathds{D}_{y}^2 \Big]\\
&=\Big[ 2 \Big( \mathds{D}_{x} + \mathds{D}_{y} \Big) \mathds{D}_{z^\prime} - \mathds{D}_{x}^2 -\mathds{D}_{y}^2 \Big].
\end{align}
Note that already the first line of the above operations would not hold if implemented with the naive finite difference operator. The treatment of the $T_6$ term requires our operator to fulfill an additional property. When the $z^\prime$ derivative after summation by parts, acts on $\delta(x-y)F_3(\frac{x+y}{2})$ we need it to affect only $F_3$. Inspecting \cref{eq:indepofzprime} we conclude that the RN-SBP operator indeed allows us to carry out this operation 
\begin{align}
    \mathds{D}_{z^\prime} \Big ( \delta(z) \circ F_3\big(\frac{z^\prime}{2}\big) \Big) = \delta(z)\circ \mathds{D}_{z^\prime} F_3\big(\frac{z^\prime}{2}\big).
\end{align}
As an example, let us replicate the manipulations in Eq. \eqref{eq:termCmanip} discretely:
\begin{equation*}
\begin{aligned}
    T_{6} & \approx \left(\delta(z) \circ F_3\left(\frac{z'}{2}\right)
    ,2(\mathds{D}_x + \mathds{D}_y )\mathds{D}_{z'} \rho\right)_\mathds{H}
    \\
    & = 
    -\left((\mathds{D}_x + \mathds{D}_y )\delta(z) \circ F_3\left(\frac{z'}{2}\right)
    ,2\mathds{D}_{z'} \rho\right)_\mathds{H}
    \\
    & = 
    -\left(\delta(z) \circ 2\mathds{D}_{z'}F_3\left(\frac{z'}{2}\right)
    ,2\mathds{D}_{z'} \rho\right)_\mathds{H}
    \\
    & = 
    -\left(\delta(z) \circ 2\mathds{D}_{z'}F_3\left(\frac{z'}{2}\right)
    ,\mathds{D}_x\rho\right)_\mathds{H}
    - \left(\delta(z) \circ 2\mathds{D}_{z'}F_3\left(\frac{z'}{2}\right)
    ,\mathds{D}_y\rho\right)_\mathds{H}
    \, .
\end{aligned}
\end{equation*}
We have thus shown that the RN-SBP operator is able to mimic all manipulations that were required in continuum to prove the conservation of the trace of the reduced density matrix.

\section{Simulating trace preserving dissipative dynamics of heavy quarkonium at high temperature}
\label{sec:NumRes}

In the following we will implement \cref{eq:mastereq} for one-dimensional $x$ and $y$, i.e. the function $\rho$ at each time step corresponds to a two dimensional array of complex numbers. To compute its time evolution we deploy the unconditionally stable and unitary Crank-Nicolson prescription \cite{crank_nicolson_1947}. 
This approach is costlier than e.g. Runge-Kutta schemes of the same order, as it involves the solution of a linear system of equations at each time step. The Crank-Nicolson method however preserves the hermiticity and positivity of the density matrix, guaranteed by the continuum Lindblad formalism. Together with the RN-SBP operator it also preserves the trace, as we will show below. The code for this study is available via a creative-commons open access attribution license at the Zenodo repository \cite{CodeZenodo}.
                
The physical quantity of interest to read off from the simulation is the survival probability of individual quarkonium quantum states as they interact with the surrounding medium. To this end we initialize the simulations with either the ground or the first excited state, corresponding to the lowest lying $\phi_0(x)$ or next to lowest eigenvector $\phi_1(x)$ of the Hamiltonian $H_S\phi_i = E_i \phi_i$ with $H_S=p^2/M+V(x)$, where $M$ denotes the heavy quark mass and $V(x)$ a real-valued interaction potential. At each step in the time evolution, we may express the density matrix in the  basis of these eigenvectors $\rho_{mn}(t)=\int dx \int dy \phi_m^*(x) \rho(x,y,t) \phi_n(y)$. The survival probability is then read off from the diagonal entries e.g. $P_0(t)=\rho_{00}(t)$.    

\cref{eq:mastereq} has been solved approximately in a previous study \cite{Miura:2019ssi} using a stochastic unravelling of the master equation in terms of an ensemble of wave functions evolving under a non-linear stochastic Schr\"odinger equation. As this so called quantum state diffusion approach necessitated additional approximations in the heavy quark velocity, the present study will provide an important crosscheck of the validity of that computation. In \cite{Kajimoto:2017rel} the master equation on the other hand has been solved through stochastic unraveling in the recoilless limit, which in the language of \cref{eq:mastereq} amounts to neglecting all contributions coming from 
the terms except for the only $D(x)$ in $F_1(x)$. 

Let us remark that standard methods of operator splitting, highly efficient for the solution of the regular Schr\"odinger equation, fail for \cref{eq:mastereq}, as we are faced with a partial differential equation including variable coefficients. The spatial dependence of the different $F$ terms leads to significant contributions of commutators in the Trotter decomposition, which break the naive counting of the Strang scheme \cite{macnamara2016operator}. In order to implement the left hand side of the master equation efficiently we therefore deploy the PETSC \cite{petsc-web-page,petsc-user-ref} sparse and distributed matrix library. 

Similarly to the values chosen in \cite{Miura:2019ssi} we discretize the density matrix on a grid of $N=256$ points in $x$ and $y$ direction each with periodic boundary conditions. Using the mass $M$ of the heavy quark to express all dimension full quantities, we have for the spatial spacing $\Delta=1/M$ and a time step of $\Delta t=0.1M(\Delta)^2$. In the simulations we will explicitly set $M=1$, from which follows $\Delta t= 0.1\Delta$. (We have checked that reducing the time step $\Delta t$ further does not significantly change the outcome of our simulations.) The interactions among the quark--anti-quark pair are captured in a model potential and dissipation kernel
\begin{align}
V(x)=-\frac{\alpha}{\sqrt{x^2+x_r^2}}e^{-m_D|x|}, \quad D(x)=\gamma e^{-x^2/\ell_{\rm corr}^2},\label{eq:parmsim2}
\end{align}
inspired by the results from high temperature perturbation theory in (3+1)d QCD. As parameter values we choose $\alpha=0.3$, $m_D=2T$, $\ell_{\rm corr}=1/T$, $\gamma=T/\pi$ and as regularization for the Coulomb potential $x_r=1/M$. The simulation will be performed for three different temperatures, $T=0.05M$, $T=0.1M$ and $T=0.3M$. Note that since the the density matrix $\rho(x,y)$ needs to fulfill periodic boundary conditions in each variable $x,y$ independently we are lead to an additional constraint for the functions $D(x)$. I.e. if $\rho(x+L,y)=\rho(x,y+L)=\rho(x,y)$ then $D(x+L/2)=D(x)$, which tells us that the function $D$ must be periodic over half the box size.

In order to carry out the Crank-Nicolson step for \cref{eq:mastereq} we have to solve a linear system of equations. Exploiting the sparse nature of the update matrix we choose to utilize the distributed sparse matrix format provided by the PETSC library. Contrary to dense matrix algorithms here the solution is found iteratively and thus approximately via the GMRES algorithm. As a compromise between precision and computational speed we select a solution tolerance of $\Delta_{\rm GMRES}=10^{-14}$ and a maximum number of steps in the iterations $N_{\rm GMRES}=100$. The error introduced by the approximation to the true solution was the dominant source of error in our simulations.

\begin{figure}
\centering
\includegraphics[scale=0.5]{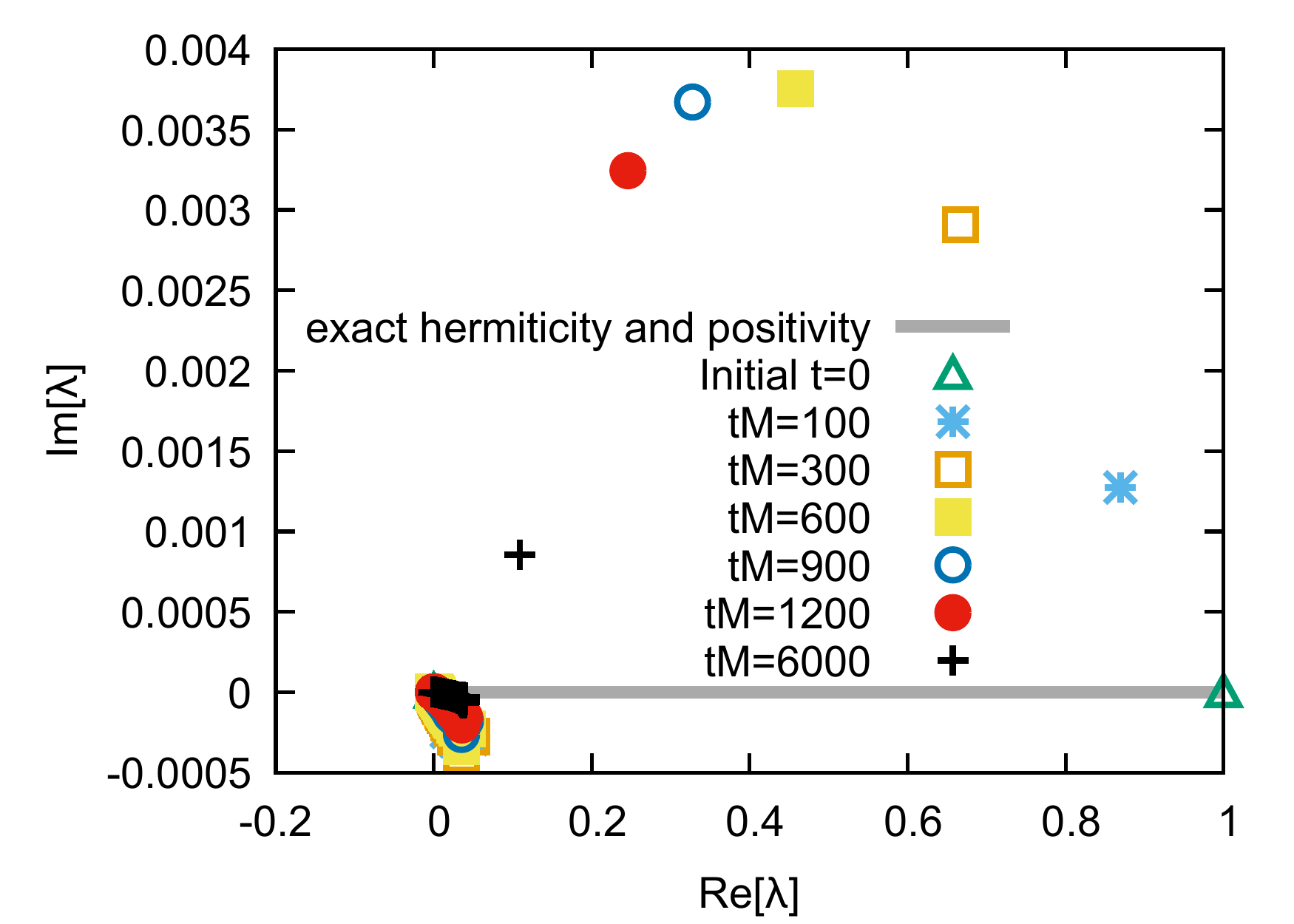}
\caption{Representative example at $T=0.1M$ of the real- and imaginary part of the eigenvalues of the matrix $\rho$ at initial time $t=0$ (open triangle) and at several later times up to $tM=6000$. Our Crank-Nicolson and RN-SBP operator based time stepping preserves the positivity of the real-part of the eigenvalues, while deviations from exact hermiticity (gray line), manifest in a finite imaginary part, are visible on the permille level. We have checked that hermiticity and positivity remain virtually unaffected by replacing the RN-SBP derivative operator by its naive counterpart.}\label{fig:eigenvalues}
\end{figure}

\begin{figure}
\centering
\includegraphics[scale=0.5]{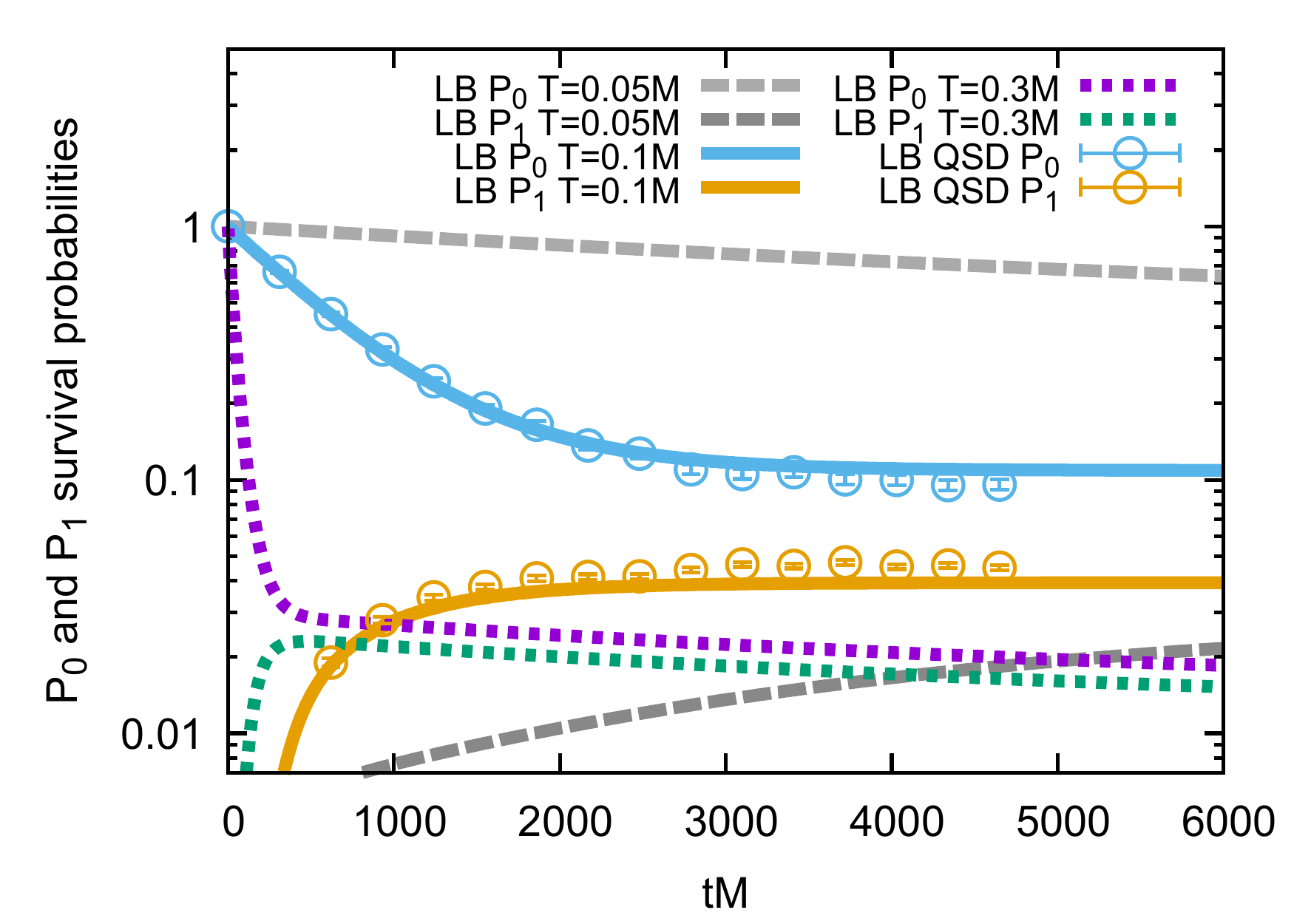}
\caption{Comparisons of the direct solution of the master equation $\eqref{eq:mastereq}$ at $T=0.1M$ (solid lines) to the approximate solution via the stochastic quantum state diffusion unraveling (open circles). The full Lindblad dynamics for $T=0.05M$ (gray dashed) and $T=0.3$ (colored dashed) are also shown.}\label{fig:cmplbqsd}
\end{figure}

Let us start by investigating the defining properties of the density matrix for the representative example of $T=0.1M$. In \cref{fig:eigenvalues} we plot the real- and imaginary part of the eigenvalues of the discretized density matrix at different times during the evolution (colored data points). The values were obtained using the SLEPC distributed eigensolver library \cite{Hernandez:2005:SSF}. We have chosen the time window such, that at the latest time $tM=6000$ the system is very close to a stationary state, which leaves the survival probabilities of the two lowest lying states of interest unchanged (see also \cref{fig:cmplbqsd}). As expected from a density matrix initialized using a single normalized eigenstate of the system Hamiltonian, it contains a single non-vanishing eigenvalue of value unity while the rest of the eigenvalues vanishes. During the time evolution, the real-part of the eigenvalues remains positive, which is a manifestation of the conservation of positivity of $\rho$. On the other hand we clearly see deviations of the imaginary part of the eigenvalues into the complex plane, which amounts to a violation of the exact hermiticity (denoted by the gray line) of the density matrix. The deviations for our choice of rather large time steps of $\Delta t=0.1/M$ however remain at the permille level at all times. We have checked that replacing the RN-SBP derivative operator with its naive counterpart leaves the positivity and hermiticity properties of the time evolution virtually unchanged.

\begin{figure}
\centering
\includegraphics[scale=0.5]{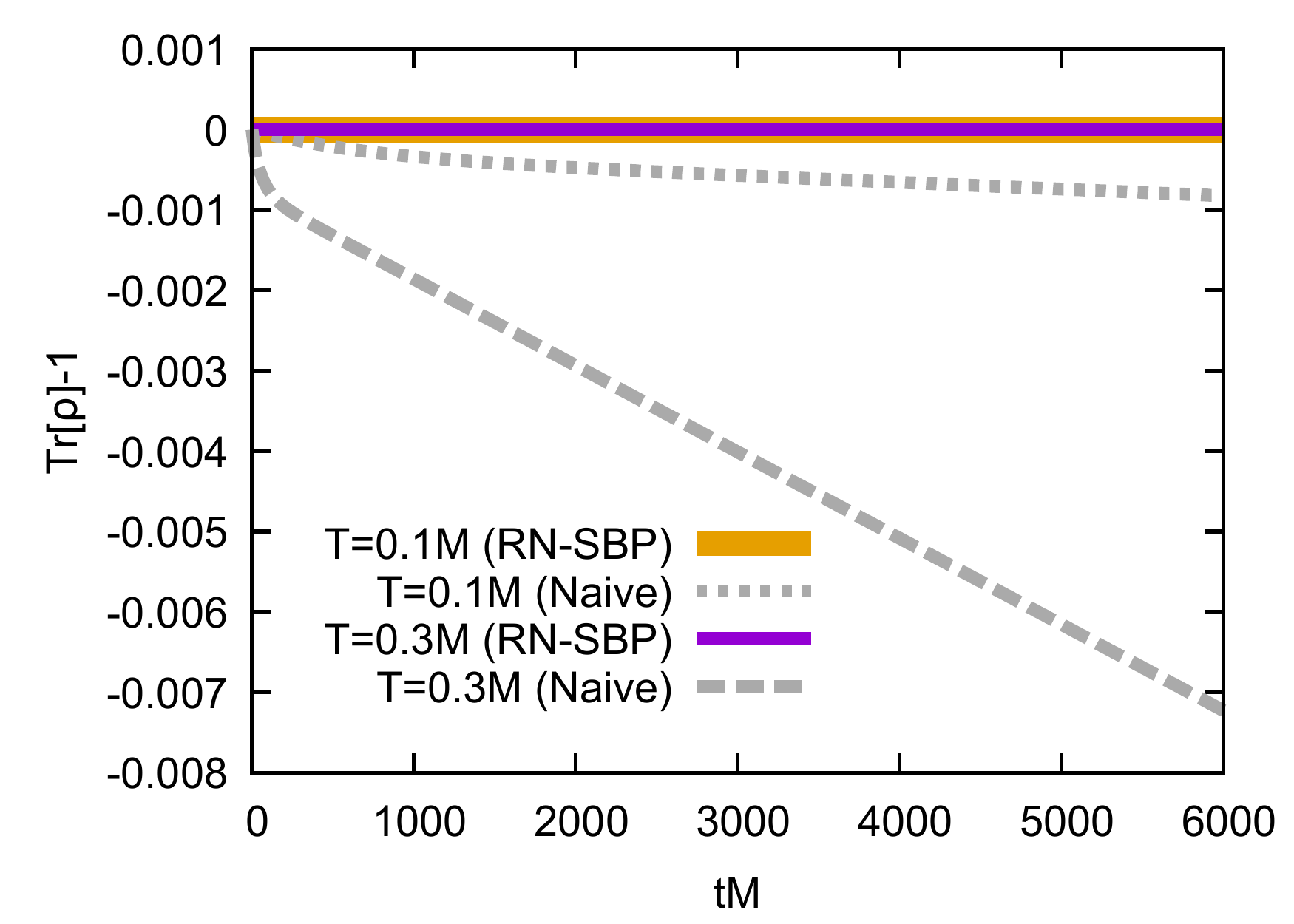}
\caption{Time dependence of the trace of the density matrix computed with the novel RN-SBP derivative operator (solid line) as well as with its naive counterpart (dashed lines) for $T=0.1M$ and $T=0.3M$. Only when we deploy the RN-SBP operator the trace is preserved excellently with deviations from unity on the level of $10^{-9}$, which are dominated by our choice of $\Delta_{\rm GMRES}=10^{-14}$. The violation of the unit trace property in case of a naive derivative operator also shows a clear dependence on the choice of parameters of the dynamical evolution.}\label{fig:trace}
\end{figure}

We continue with an inspection of the trace of the density matrix, based on the novel RN-SBP derivative operator and plotted as solid lines in \cref{fig:trace} for $T=0.1M$ and $T=0.3M$. Due to the properties of the RN-SBP operator derived in \cref{sec:RNSBPOp}, we find that the trace values are excellently preserved with a maximum deviation from unity of $10^{-9}$. For comparison purposes we also plot the values of the trace as obtained with the naive difference operator as dashed lines. One can clearly see that the violation of the unit trace property grows with time and that the strength of the deviation depends on the particular parameters of the dynamical evolution. We have checked that the minute deviation from unit trace in case of the RN-SBP operator reduces when we lower the tolerance $\Delta_{\rm GMRES}$ for the iterative solution of the the Crank-Nicolson step. Reducing the tolerance further, at some point the errors introduced by the finite $\Delta_{\rm GMRES}$ are no longer the dominant source of error and instead it is the finiteness of the time step $\Delta t$. At this point both $\Delta t$ and $\Delta_{\rm GMRES}$ need to be reduced in tandem for the results to further improve.

\begin{figure}
\centering
\includegraphics[scale=0.5]{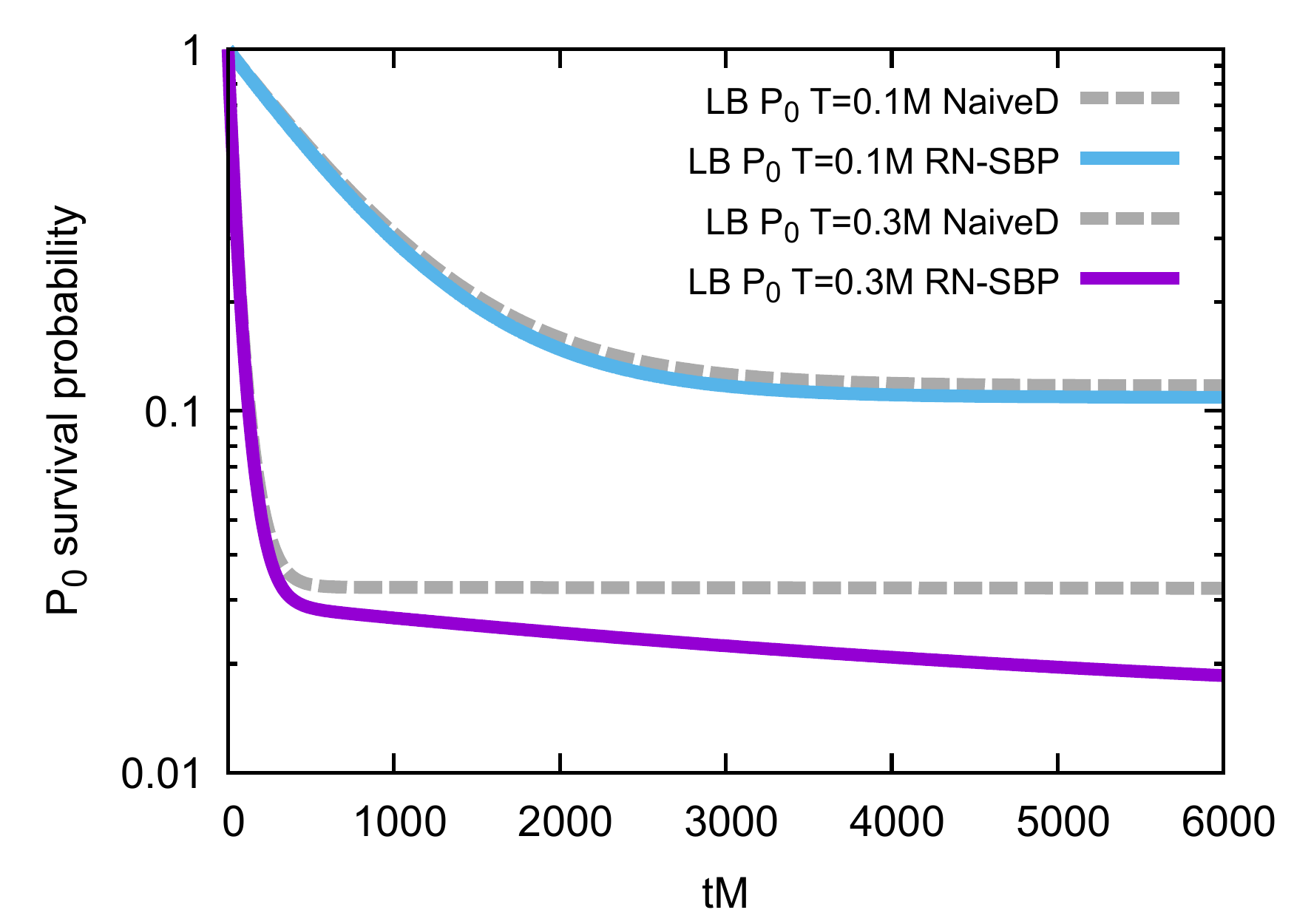}
\caption{Comparison of the ground state survival probabilities $P_0$ at $T=0.1M$ and $T=0.3M$ based on the novel RN-SBP difference operator (solid lines) and the corresponding values using the naive derivative operator (dashed lines). While for $T=0.1M$ the difference at $tM=6000$ is around 8\% it already grows to 57\% for $T=0.3M$.}\label{fig:cmpRNSBPNaive}
\end{figure}

One may question whether an apparently small deviation from unit trace by less than a percent, as is visible in \cref{fig:trace}, has any significant consequences for the physics outcome of our simulation. To this end we compare the simulated values of the ground state survival probability $P_0$ in \cref{fig:cmpRNSBPNaive} using the novel RN-SBP derivative operator (solid lines) and the naive counterpart (dashed lines) at $T=0.1M$ and $T=0.3M$. Already for $T=0.1M$, where the maximum trace deviation was below one permille, at late times $tM=6000$, we find a disagreement of around 8\%. The difference becomes even more pronounced for $T=0.3M$, where a trace deviation of around seven permille translates into a disagreement of the survival probabilities of 57\%.

We conclude, based on the above comparison, that that the combination of the Crank-Nicolson scheme with our novel RN-SBP operator provides an accurate discrete representation of the dynamics described by \cref{eq:mastereq}.

Having convinced us of the inner workings of the underlying discretization, we proceed to investigate the physics results of our simulation. In \cref{fig:cmplbqsd} we plot the survival probabilities of the ground and first excited states of the system Hamiltonian $P_0$ and $P_1$ for the case of $T=0.05M$ as gray dashed lines, for $T=0.1M$ as colored solid lines, and for $T=0.3M$ as colored dashed lines. As crosscheck of our previous work we also plot the values obtained from an approximate stochastic unravelling of the master equation via the quantum-state-diffusion approach as open circles.

We find that the stochastic unraveling provides a very good description of the dynamics of the ground state up to $tM=3000$, which is already longer than what would be needed in the simulation of a realistic heavy-ion collision. At late times deviations from the direct solution of the master equation become visible, which however remain at the 20\% level. The first excited state shows similar deviations, which set in at a bit earlier times around $tM=1250$. 

The simulation of the evolution of the quarkonium system at different temperatures proceeds in a consistent fashion. In a colder environment, such as $T=0.05M$ (gray dashed lines) the medium is unable to interfere with the quarkonium binding as efficiently as is the case at $T=0.1M$. This on the one hand leads to a slower decay of the ground state survival probability and on the other hand produces a less rapid population of the excited states. Conversely in a hotter environment, such as at $T=0.3M$ (colored dashed lines) the ground state is more efficiently depleted while rapidly populating the excite states.

The relative abundances between the states in thermal equilibrium are expected to be governed by the Boltzmann distribution, which means that the two curves will lie further apart at $T=0.05M$ and closer together at $T=0.3M$ than at $T=0.1M$. As we see in \cref{fig:cmplbqsd}, at first rising temperatures lead to stronger occupation of the exited states but eventually at high enough temperatures also their contribution will become suppressed.

We see that a steady state is reached at different times for different temperatures. Comparing $T=0.05M$ and $T=0.1M$ we see that at higher temperature the steady state emerges already at $tM=3000$, while at the lower temperature we need to wait until around $tM=30000$. Interestingly for $T=0.3M$ we find that the relative abundances between the ground and first excited state are established quite quickly, around $Mt=1500$ but that the overall amplitude of the survival decreases over time, indicating that the excited states are not yet equilibrated\footnote{Note that at $T=0.3M$ the eigenstates of the Hamiltonian only contain a single bound state and thus the behavior of the majority of the lowest lying states is affected by the finite volume of the simulation}.

\begin{figure}
\centering
\includegraphics[scale=0.5]{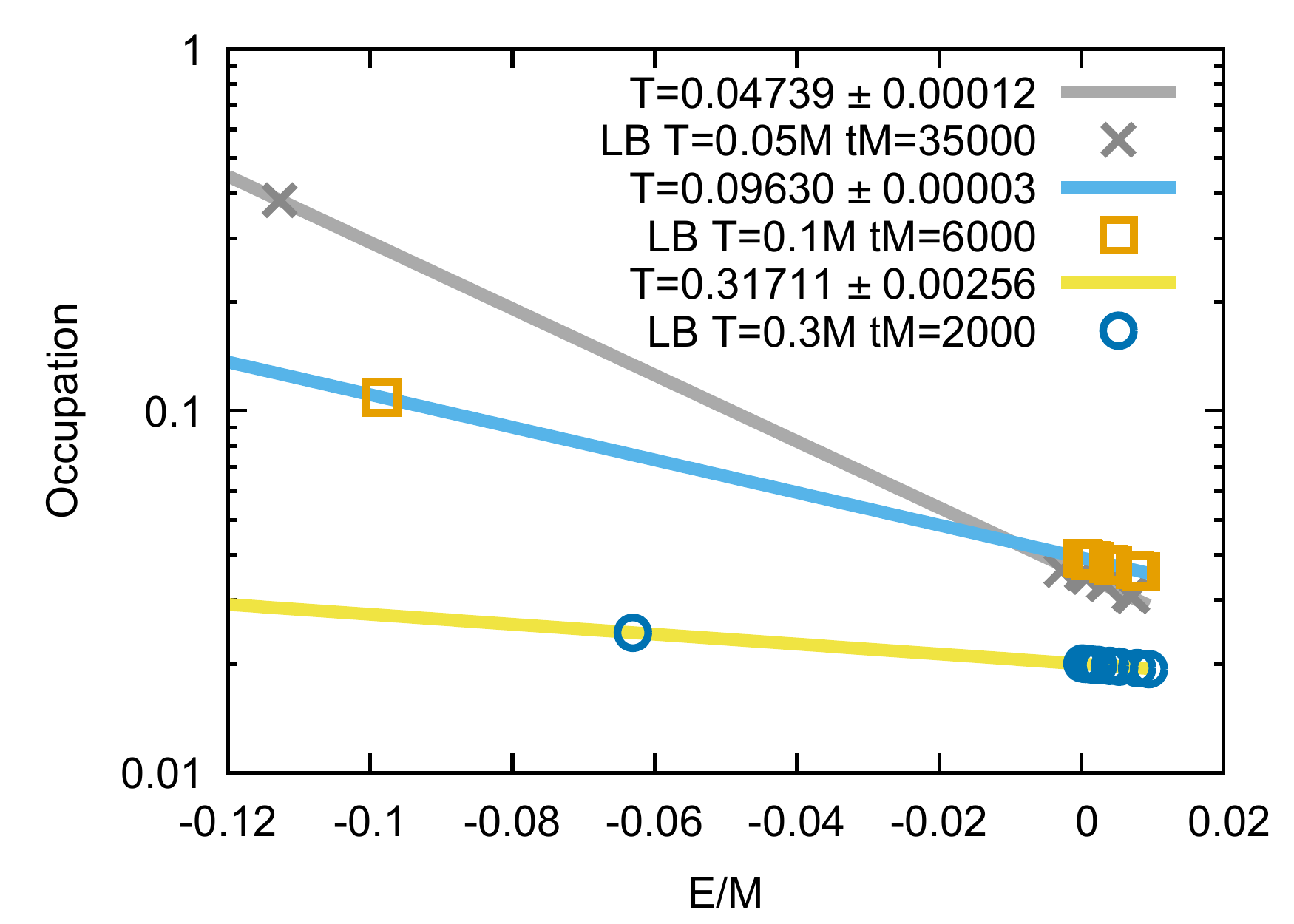}
\caption{Comparison of the abundances of the individual states in form of the survival probabilities $P_i$ versus the energy of these states (data points). As solid lines we plot exponential fits, motivated by the expected Boltzmann distribution, with the corresponding best fit temperature given in the key.}\label{fig:spectra}
\end{figure}

The properties of the steady state reached at late times may be investigated by inspection of the abundances of the individual states present in the system. In \cref{fig:spectra} we show the corresponding values of the survival probabilities $P_i$ vs. the energy of the states as individual data points. Motivated by the expectation that eventually a Boltzmann distribution will emerge, we also plot exponential fits to the data and provide the best fit value of the "inverse slope parameter", the temperature, in the key. For all systems at $T=0.05M$, $T=0.1M$ and $T=0.3M$ we find that the fit captures all of the ten lowest lying states very well and a temperature emerges, which, while not exactly at the environment, lies very close to it. Note that as was shown in \cite{Miura:2019ssi} such a deviation is not unexpected, as thermalization with the same temperature, can only be proven in the classical limit and for small velocities, quantum corrections may lead to small deviations.

\begin{figure}
\centering
\includegraphics[scale=0.38]{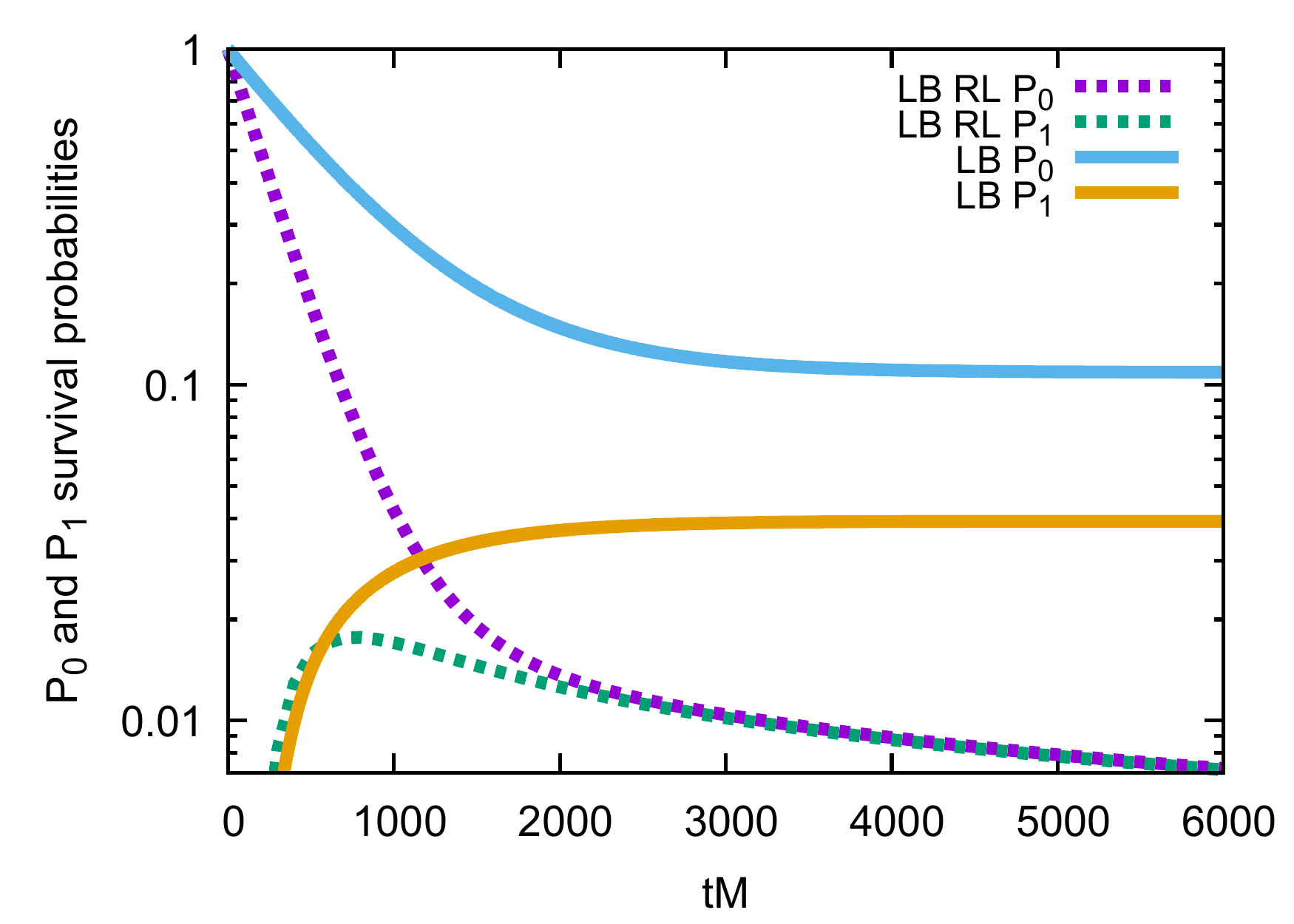}
\includegraphics[scale=0.38]{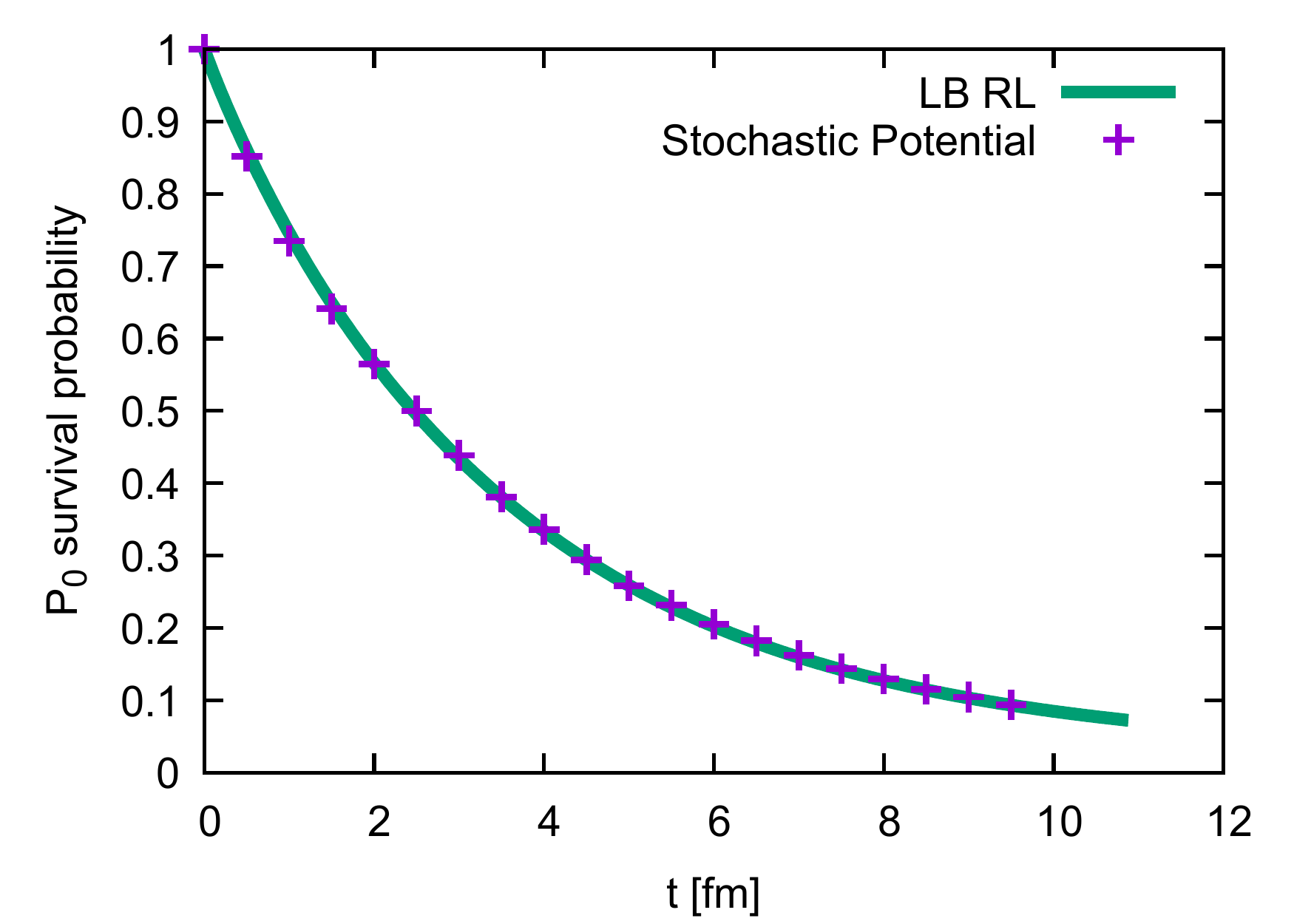}
\caption{(left) Comparions of the full dissipative dynamics of \cref{eq:mastereq} at $T=0.1M$ (solid lines) to the recoilless limit (dashed lines) at the same temperature. (right) Crosscheck of the dynamics in the recoilless limit, computed from the Lindblad equation (solid line) to the values obtained via the stochastic potential using the paramters of \cite{Kajimoto:2017rel}. }\label{fig:cmplbsp}
\end{figure}

In previous studies, such as Refs.~\cite{Akamatsu:2011se,Rothkopf:2013kya,Kajimoto:2017rel}, the quarkonium dynamics were investigated in the recoilless limit, which allows \cref{eq:mastereq} to be unraveled in terms of a stochastic potential. It amounts to neglecting all ${\bm F}_2$ and $F_3^{ij}$ terms and retaining only the $D(x)$ term in $F_1$. For illustrative purposes let us compare the full dissipative dynamics to this approximation in the left panel of \cref{fig:cmplbsp}, similar to a comparison previously carried out in \cite{Miura:2019ssi}. As dissipative effects are absent, the fluctuations of the environment transfer energy into the quarkonium system which is unable to release it back to the medium. Hence the system heats up unabated and one expects that eventually all states will become destabilized. And indeed the dashed lines clearly show this behavior, as the survival probabilities $P_0$ and $P_1$ eventually fall on top of each other. We also emphasize that already at early times a significant deviation from the full dissipative dynamics is observed for the ground state. As a crosscheck we show in the right panel of \cref{fig:cmplbsp} the comparison of the dynamics in the recoilless limit obtained from the solution of the approximate master equation (solid line) and via the stochastic potential approach (data points) using here as an exception the parameters of \cite{Kajimoto:2017rel}, confirming the correctness of the numerics of that study.

\begin{figure}
\centering
\includegraphics[scale=0.5]{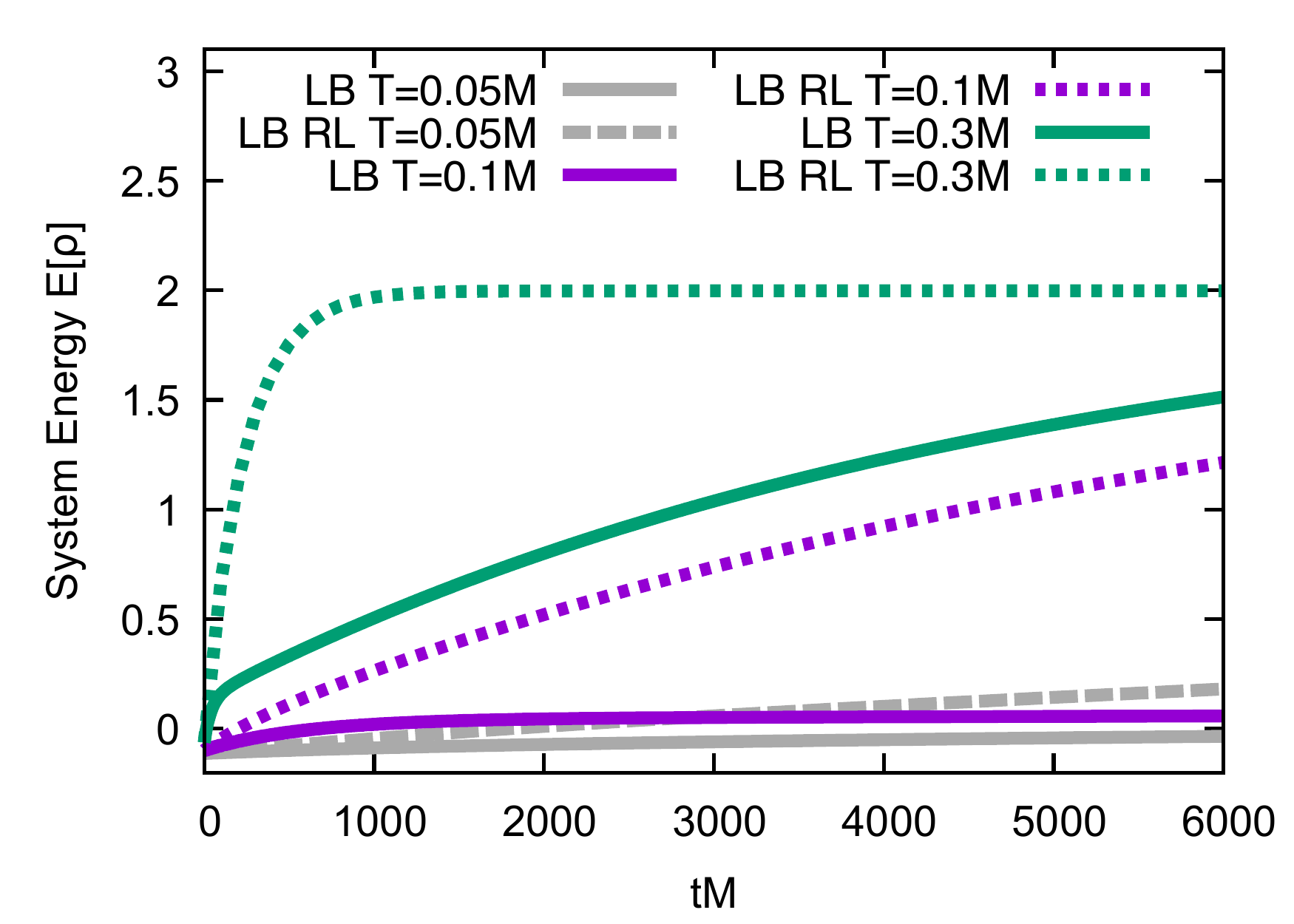}
\caption{Comparison of the total system energy of the system evolving according to fully dissipative dynamics (solid lines) and in the recoilless limit (dashed lines) at the temperatures $=0.05M$, $T=0.1M$ and $T=0.3M$ (consecutively higher lying lines).}\label{fig:energy}
\end{figure}

The failure of the recoilless limit to thermalize also manifests itself clearly in the total energy $E={\rm Tr}[\hat H \hat \rho ]$ of the system, as shown in \cref{fig:energy}. The solid lines represent the fully dissipative dynamics, which asymptote against a constant value at late times. On the other hand the dissipationless dynamics depicted via dashed lines for $T=0.1M$ and $T=0.05M$ exhibit an unabated rise that is linear in time at late times. At $T=0.3M$ the energy in the dissipationless limit eventually runs into a constant too, which is not related to thermalization, but to reaching the "infinite temperature limit" on a lattice with finite extent and lattice spacing.

\section{Conclusion and Outlook}
\label{sec:conclout}

In this study, we have presented an improved numerical open quantum systems treatment of heavy quarkonium at high temperature, via its Lindblad equation. We showed that in order to fulfill the defining properties of the quarkonium density matrix $\rho(x,y)$, guaranteed by the Lindblad equation in the continuum, we need to be able to carry out integration by parts and exploit the reparametrization of the system between the $(x,y)$ and $(z=x-y,z^\prime=x+y)$ coordinates. In order for the properties to hold also after discretization we thus developed a novel RN-SBP derivative operator, which not only fulfills the summation-by-parts property, but also remains neutral under the above mentioned reparametrization.

With the novel RN-SBP operator at hand, we presented a numerical implementation of \cref{eq:mastereq} using the Crank-Nicolson approach. It allowed us to evolve the density matrix in time, while preserving its positivity, hermiticity and trace accurately. In turn we were able to not only crosscheck the validity of previous computations, based on the approximate Quantum State Diffusion approach, but also obtain a more robust result for the steady state occupancies at late times. 

There are several directions to explore next: on the one hand it will be interesting to formulate the novel RN-SBP operator for use in higher dimensions, as the ultimate goal is to solve Lindblad equations, such as \cref{eq:mastereq} for three-dimensional coordinate ${\bm x}$ and ${\bm y}$. Given its simple structure in one dimension, its generalization via shifts along different axes appears straight forward, but has to be verified explicitly. In addition one may want to consider how to formulate compatible second order derivatives, which require less off-diagonal terms, compared to the naive application of the first order derivative operator twice. In order for RN-SBP operators to play out their full strength in precision studies of dissipative dynamics, also higher order incarnations of the first derivative operator need to be formulated.

In summary the novel RN-SBP operator presented here provides an interesting discrete implementation of a continuum derivative property not treated explicitly in the literature so far. In turn, we hope that it will be of benefit in many other numerical settings, be it for the study of dissipative dynamics or beyond. 

\subsection*{Acknowledgements}

The work of Y.A. is supported by JSPS KAKENHI Grant Number JP18K13538. O.${\rm \mathring{A}}$., F.L. and J.N. acknowledge funding from the Swedish Research Council (Stockholm) under grant number 2018-05084\_VR and from the Swedish e-Science Research Center (SeRC). A.R. gladly acknowledges support by the Research Council of Norway under the FRIPRO Young Research Talent grant 286883. This work has utilized computing resources provided by  
UNINETT Sigma2 - the National Infrastructure for High Performance Computing and Data Storage in Norway under project NN9578K-QCDrtX "Real-time dynamics of nuclear matter under extreme conditions"

\bibliographystyle{elsarticle-num}

\bibliography{TracePreservingDynSBP}

\end{document}